\documentclass[epj,nopacs,final]{svjour}
\usepackage{graphics}
\usepackage{latexsym,overpic,rotating}
\usepackage{subfigure,color,showkeys}
\usepackage{epsfig,color,rotating,amsmath,delarray,array,multirow}
\usepackage{epstopdf}
\usepackage{makeidx,pifont,float,amssymb}
\usepackage{ulem} 
\newcommand{\er}{$\pm$}
\newcommand{\be}{\begin{eqnarray}}
\newcommand{\ee}{\end{eqnarray}}
\newcommand{\bea}{\begin{eqnarray}}
\newcommand{\eea}{\end{eqnarray}}
\newcommand{\bc}{\begin{center}}
\newcommand{\ec}{\end{center}}

\newcommand{\nn}{{\nonumber}\\}
\newcommand{\beq}{\begin{equation}}
\newcommand{\eeq}{\end{equation}}
\newcommand{\ba}{\begin{eqnarray}}
\newcommand{\ea}{\nonumber \end{eqnarray}}

\newcommand{\bi}{\begin{enumerate}}
\newcommand{\ei}{\end{enumerate}}

\definecolor{darkyellow}{rgb}{1.0,0.5,0}
\definecolor{lightyellow}{cmyk}{0,0,0.5,0}
\definecolor{lightred}{rgb}{1,0.5,0.5}
\definecolor{lightgreen}{rgb}{0,0.4,0}
\definecolor{lightblue}{rgb}{0.5,0.5,1}
\definecolor{darkred}{rgb}{0.8,0,0}
\definecolor{darkgreen}{rgb}{0,0.4,0}
\definecolor{darkcyan}{cmyk}{1,0.3,0.3,0.3}
\definecolor{darkblue}{rgb}{0,0,0.6}
\definecolor{lightbrown}{rgb}{0.7,0.3,0.3}
\definecolor{darkbrown}{rgb}{0.5,0,0}
\definecolor{violett}{rgb}{0.6,0,0.8}
\begin{document}

\title{\boldmath Hyperon I: Study of the $\Lambda(1405)$}
\titlerunning{Hyperon I: Study of the $\Lambda(1405)$}

\author{A.V. Anisovich\inst{1,2}, A.V.~Sarantsev\inst{1,2}, V.A. Nikonov\inst{1,2},
V. Burkert\inst{3}, R.A. Schumacher\inst{4}, U. Thoma\inst{1}, and E.~Klempt\inst{1}\thanks{email: klempt@hiskp.uni-bonn.de}\\[-2ex]
}
\authorrunning{A.V. Anisovich {\it et al.}}

\institute{
\inst{1}Helmholtz--Institut f\"ur Strahlen-- und Kernphysik, Universit\"at Bonn, 53115 Bonn, Germany\\
\inst{2}National Research Centre ``Kurchatov Institute'', Petersburg Nuclear Physics Institute,
        Gatchina, 188300 Russia\\
\inst{3}Thomas Jefferson National Accelerator Facility, Newport News, Virginia 23606\\
\inst{4}Carnegie Mellon University, Pittsburgh, Pennsylvania 15213, USA
 }

\date{\today}

\abstract{Low-energy data on the three charge states in $\gamma p \to K^+(\Sigma\pi)$ from CLAS at
JLab, on $K^-p\to \pi^0\pi^0\Lambda$ and $\pi^0\pi^0\Sigma$ from the Crystal Ball at BNL, bubble
chamber data on $K^-p\to\pi^-\pi^+\pi^{\pm}\Sigma^{\mp}$, low-energy total cross sections on $K^-$
induced reactions, and data on the $K^-p$ atom are fitted with the BnGa partial-wave-analysis
program. We find that the data can be fitted well with just one isoscalar spin-1/2 negative-parity
pole, the $\Lambda(1405)$, and background contributions. }

\maketitle

\section{Introduction}
The $\Lambda(1405)1/2^-$ resonance -- here written as $\Lambda(1405)$ -- has been discussed
controversially since its discovery in 1961~\cite{Alston:1961zzd}. Dalitz and collaborators
considered the $\Lambda(1405)$ as a quasibound molecular state of the $\bar{K}N$ system
\cite{Dalitz:1960du,Dalitz:1967fp}. In quark models, this resonance is interpreted as $qqq$
resonance in which one of the quarks is excited to the $p$ state; jointly with its spin partner
$\Lambda(1520)3/2^-$ it forms a spin-doublet SU(3)-singlet, as expected within
SU(6)~\cite{Isgur:1978xj}. Later, Kaiser, Waas and Weise constructed an effective potential from a
chiral Lagrangian, and the $\Lambda(1405)$ emerged as quasi-bound state in the $\bar KN$ and
$\pi\Sigma$ coupled-channel system \cite{Kaiser:1996js}. Oller and Meissner~\cite{Oller:2000fj}
studied the $S$-wave $\bar K N$ interactions in a relativistic chiral unitary approach based on a
chiral Lagrangian. The Lagrangian was obtained from the interaction of the SU(3) octet of
pseudoscalar mesons and the SU(3) octet of stable baryons. In their coupled-channel approach, they
found two isoscalar resonances below 1450\,MeV, at 1379.2\,MeV and at 1433.7\,MeV, and one
isovector resonance at 1444.0\,MeV. The authors of Ref.~\cite{Jido:2003cb} suggested that the two
$\Lambda^*$ poles as well as a third state at 1680\,MeV are combinations of the singlet state and the
two octet states expected in the  $8\otimes8$ into $1\oplus8_s \oplus 8_a
\oplus10\oplus \overline{10}\oplus 27$. They interpreted the first wider state (at 1390\,MeV in
their analysis) as mainly singlet, a second and a third state at 1426\,MeV and 1680\,MeV as
mainly octet states. The isovector sector was found to be much more sensitive on the details of the
coupled channel approach~\cite{Jido:2003cb}. Based on the approach used in~\cite{Oller:2000fj}, two
poles were found at 1401\,MeV and 1488\,MeV~\cite{Jido:2003cb}, based on~\cite{Oset:2001cn}, one
state was found at 1580\,MeV. Here the other isovector state disappeared for dynamical reasons. The
$\Sigma$ resonances were interpreted as isovector companions of the isoscalar states. The findings
presented in~\cite{Oller:2000fj} and~\cite{Jido:2003cb} were confirmed in a number of further
studies. Here we quote a few recent
papers~\cite{Cieply:2009ea,Ikeda:2012au,Guo:2012vv,Mai:2012dt,Mai:2014xna,Roca:2013av,Roca:2013cca,Miyahara:2018onh,Feijoo:2018den}.
A survey of the literature and a discussion of the different approaches can be found in
Ref.~\cite{Cieply:2016jby}.

In quark models~\cite{Capstick:1986bm,Glozman:1995fu,Loring:2001ky,Giannini:2015zia}, three
isoscalar resonances are expected below 1.9\,GeV. $\Lambda(1405)$ is interpreted as the (mainly)
SU(3) singlet state. The  four-star $\Lambda(1670)$ $1/2^-$ has a mass 140\,MeV above
$N(1535)1/2^-$; the three-star $\Lambda(1800)1/2^-$ is found 150\,MeV above $N(1650)1/2^-$. These
two states are commonly identified with the two expected (mainly) octet
states~\cite{Loring:2001ky}. The $\Sigma(1620)1/2^-$ resonance is interpreted as the isospin
partner of $\Lambda(1670)1/2^-$ and $\Sigma(1750)1/2^-$ as the isospin partner of
$\Lambda(1800)1/2^-$. This interpretation is supported in a forthcoming study of the spectrum of
hyperon resonances~\cite{hyperon-II,hyperon-III}. The SU(3) symmetry of the quark model is thus
experimentally confirmed. An assignment of the two resonances at 1426\,MeV and 1680\,MeV to the
quark model SU(3) octet states with spin-parity $1/2^-$, instead of the $\Lambda(1670)$ and
$\Lambda(1800)1/2^-$, would be at variance with the quark model.

Many, possibly all, dynamically generated baryon resonances like $N(1440)1/2^+$, $N(1535)1/2^-$,
$\Delta(1700)3/2^-$, $\cdots$ can be map\-ped onto the spectrum expected in quark models,
except of course the pentaquark candidates $P_c(4380)$ and $P_c(4450)$~\cite{Aaij:2015tga}.
With the
identification of the negative-parity $\Lambda$ resonances as outlined above, one of the two
low-mass $\Lambda$ states and the low-mass $\Sigma$ state
in~\cite{Jido:2003cb,Oset:2001cn} cannot be mapped
onto quark-model states: the two states are supernumerous (and not required in the analysis
presented here). Based on Regge phenomenology, the authors of Ref. \cite{Fernandez-Ramirez:2015fbq}
argue that the narrow state at about 1430\,MeV fits into the common pattern of a linear Regge
trajectory of known three-quark hyperons possibly indicating its three-quark nature. The wider
state below $\approx$1400\,MeV is speculated to be a pentaquark or of molecular nature.

The two-pole structure of the $\Lambda(1405)$ region is not uncontested. All work before
\cite{Oller:2000fj} assumed a single pole in this region. Later, HADES data on the reaction $p +
p\to \Sigma^+ +\pi^- +K^+ +p$  were successfully fitted with a single
$\Lambda(1405)$~\cite{Agakishiev:2012xk}, and this result was confirmed in a subsequent
reanalysis~\cite{Hassanvand:2012dn}. The CLAS collaboration studied the three charge states in the
reaction $\gamma p\to K^+\Sigma\pi$~\cite{Moriya:2013eb} which provide precise information on the
$\Lambda(1405)$ line shape. Its spin and parity were determined in \cite{Moriya:2014kpv},  until
then taken from the quark model. The data were fitted in~\cite{Moriya:2013eb}, the best fit was
achieved with two low-mass isovector states ($\Sigma^*$'s) and one  isoscalar state
$\Lambda(1405)$. A reanalysis of these data showed that the data are also compatible with a
standard single-pole $\Lambda(1405)$ \cite{Hassanvand:2017iif}. Dong, Sun and
Pang~\cite{Dong:2016auh} solved the Bethe-Salpeter equation in an unitary coupled-channel ansatz
taking relativistic effects and off-shell corrections into account. In their model, the authors found that the off-shell
corrections are very important. Without these, the authors reproduced the two-pole structure. Yet one pole
disappeared when the off-shell corrections were switched on, and only one $\Lambda(1405)$ survived. This contradicts~\cite{Mai:2012dt,Bruns:2010sv}; in their ansatz, off-shell effects were found to be small and two poles were present. 
Myint {\it al.}~\cite{Myint:2018ypc} used a chiral model and found two poles in the $\Lambda(1405)$
region. The peak structure in the data was assigned to a single pole while the second one provided
a continuum background amplitude affecting the shape of the peak, but that pole was not interpreted
as genuine resonance.

Direct experimental evidence for the presence of two poles in the $\Lambda(1405)$ region has been
reported~\cite{Lu:2013nza}. The CLAS collaboration studied electroproduction of this resonance by
studying the reaction $e^-p \to e^- K^+(p\pi^0)\pi^-$ with the $p\pi^0$ mass being compatible with
$\Sigma^+$ and with four-momentum transfers ranging from $-t=0.5$ to 4.5 GeV$^2$. The data were
shown for two subsets with $1.0<Q^2<1.5$GeV$^2$ and $1.5<Q^2<3.0$\,GeV$^2$. The latter data were
fitted with two incoherent Breit-Wigner functions with $\Sigma\pi$ as only decay channel. The
masses optimized at 1.368\er0.004\,GeV and 1.423\er0.002\,GeV (statistical fit errors only).  A
possible $\Sigma(1385)3/2^+$ contribution was estimated to be small. The low-$t$ data set was not
fitted simultaneously, and seem not describable with the same assumptions. Also the related chain
$e^-p \to e^- K^+(p\pi^-)\pi^0$ -- which avoids possible $\Sigma(1385)3/2^+$ contaminations -- has
not been investigated.

In this paper we present a partial wave analysis of data covering the $\Lambda(1405)$ region. The
data include the low-mass part of the $\Sigma\pi$ system in the reaction $\gamma p\to K^+\Sigma\pi$
from JLab~\cite{Moriya:2013eb}, data on the reaction $K^-p\to\pi^0\pi^0\Sigma^0$ from BNL
\cite{Prakhov:2004an} and bubble chamber data on $K^-p\to\pi^-\pi^+\pi^{\pm}\Sigma^{\mp}$
\cite{Hemingway:1984pz}, differential cross sections for $K^-p\to K^-p$ and $K^-p \to\bar K^0n$
from \cite{Mast:1975pv}, total cross section
measurements~\cite{Humphrey:1962zz,Watson:1963zz,Sakitt:1965kh,Ciborowski:1982et}, ratios of $K^-p$
capture rates~\cite{Tovee:1971ga,Nowak:1978au}, and the recent experimental results on the energy
shift and width of kaonic hydrogen atoms which constrain the $K^-p$ $S$-wave scattering length
\cite{Bazzi:2011zj,Bazzi:2012eq}. Within the BnGa ansatz, the data are fully compatible with just
one isoscalar resonance and conveniently chosen background amplitudes.

\section{\label{PWA}Formalism}

In this section the basic features of the dispersion integration method are considered for the
scattering amplitude. We start from the $K$-matrix method. This approximation extracts the leading
singularities, it is a very popular approach in partial wave analyses. The pole and threshold
singularities of the partial wave amplitude are taken into account, and the amplitude automatically
satisfies the unitarity condition. Here we describe the dynamical amplitude without the angular
momentum tensors needed for non-vanishing angular momenta. The full amplitude is discussed 
in Ref.~\cite{hyperon-II}.

Although the K-matrix amplitude is an analytic function in the complex plane, it neglects left-hand
singularities of the partial wave amplitude. Near thresholds, the $K$-matrix approach generates
false kinematical singularities that need to be suppressed by imposing new assumptions. As a
result, the $K$-matrix approach is not reliable in the low energy region: this was clearly
demonstrated in the analysis of the $\pi\pi$ S-wave scattering amplitude near the $\pi\pi$
threshold~\cite{Gasser,Caprini:2005zr}.

\subsection{Spectral integral equation for the K-matrix amplitude}

The $K$-matrix approach was introduced to satisfy directly the unitarity condition which is very
important for an analysis of the reactions near the unitarity limit. The $S$-matrix for transition
between different final states can be written as
 \be S=\left(I+i\hat \rho\hat
K\right)\left(I-i\hat \rho\hat K\right)^{-1}\hspace{-2mm}=\hspace{-0.5mm}I+2i\hat\rho\hat
K\left(I-i\hat \rho\hat K\right)^{-1}\,.
 \ee
Here, $\hat\rho$ is a diagonal matrix  describing the phase volumes and $\hat K$ is a real matrix
which describes resonant and non-resonant contributions.

For the partial wave amplitude $A(s)$ one obtains
 \be \hat A=\hat K\left(I-i\hat \rho\hat
K\right)^{-1}\hspace{-2mm}=\hspace{-0.5mm}\hat K+\hat Ki\hat\rho \hat K+\hat Ki\hat\rho \hat
Ki\hat\rho \hat K+\ldots
\label{km}
 \ee
This equation can be also rewritten as
 \be \hat A=\hat A\,i\,\hat \rho \hat K+\hat K. \label{km_a}
 \ee
The factor $(I-i\hat \rho\hat K)^{-1}$ describes the rescattering in the final state, it is
inherent not only for scattering amplitudes but for production amplitudes as well.

The elements of the K-matrix are parameterized as a sum of resonant terms (first-order poles) and
non-resonant contributions:
 \be K_{ij}= \sum_\alpha \frac{g^{(\alpha)}_i\,g^{(\alpha)}_j}
{M^2_\alpha-s}+f_{ij}\;. \label{km_el}
 \ee
This form is defined by the symmetry condition and the condition that the scattering amplitude has
pole singularities of the first order.

This approach allows us to distinguish between ``bare'' and ``dressed'' particles: due to
rescattering, the bare particles, with poles on the real-$s$ axis, are transformed into particles
dressed by a ``coat'' of mesons. In the $K$-matrix approach we deal with a ``coat'' formed by real
particles. The contribution of virtual particles is included in the main part of the loop diagram,
$B(s)$, discussed below, and is taken into account effectively by the renormalization of mass and
couplings.

Let us discuss hadron-hadron scattering and the production amplitudes using the dispersion-relation
(or spectral integral) technique. We write for the $K$-matrix amplitude a spectral integral
equation which is an analog of the Bethe-Salpeter equation~\cite{Salpeter} for the Feynman
technique. The spectral integral equation for the transition amplitude from the channel $a$ to
channel $b$ is given by
 \be
 A_{ab}(s)=\hspace{-1mm}
\int\frac{ds'}{\pi}\frac{A_{aj}(s,s')}{s'-s-i\epsilon}\rho_j(s') K_{j\, b}(s',s)+K_{ab}(s)\,.
\label{bethe}
 \ee
Here, $\rho_j(s')$ is the diagonal matrix of the phase volumes, $A_{aj}(s,s')$ the off-shell
amplitude and $K_{j\,b}(s,s')$ the off-shell elementary interaction. The term $-i\epsilon$
indicates that the integration is carried out in the complex plane just below the real axis.

The standard way of transforming Eq.~(\ref{bethe}) into a $K$-matrix form is the extraction of the
imaginary and principal parts of the integral. The principal part has no singularities in the
physical region and can be omitted (or taken into account by a renormalization of the $K$-matrix
parameters):
 \bea
 \int\frac{ds'}{\pi}\frac{A_{aj}(s,s')}{s'-s-i\epsilon}\rho_j(s')
K_{j\,b}(s',s)&=& \nn
 P\hspace{-4mm}\int\frac{ds'}{\pi}\frac{A_{aj}(s,s')}{s'-s}\rho_j(s')
K_{j\,b}(s',s)&+&iA_{aj}(s,s)\rho_j(s)K_{j\,b}(s) \nn
\to iA_{aj}(s,s)\rho_j(s)K_{j\,b}(s),
 \label{bethe2}
 \eea
where $\int\hspace{-2.5mm}p$ is the principle-value integral. We thus obtain the standard K-matrix
expression (\ref{km_a}).

One of the easiest ways to take into account the real part of the integral in Eq.~(\ref{bethe2})
(the so-called dispersion corrections) is to assume that the amplitude and the $K$-matrix have a
trivial dependence on $s'$. Such a case corresponds, e.g., to a parameterization of the resonant
couplings and non-resonant $K$-matrix terms by constants and to a regularization of the integral in
Eq.~(\ref{bethe2}) which depends on the scattering channel only by subtraction at a fixed energy.
In this case we obtain
 \bea
\int\frac{ds'}{\pi}\frac{A_{aj}(s,s')}{s'-s-i\epsilon}\rho_j(s') K_{j\,b}(s',s)=\\
A_{aj}(s,s)Re\,B(s) K_{j\,b}(s,s)+& iA_{aj}(s,s)\rho_j(s)K_{j\,b}(s) \nonumber
\label{bethe3}
 \eea
 \be
 {\rm where}\qquad\ Re\,B(s)=P\hspace{-4mm}\int\frac{ds'}{\pi}\,\frac{\rho_j(s')}{s'-s}
 \ee
And for the transition amplitude we obtain
 \bea A&=&K\left (I-\hat Re\,B\hat K-i\hat\rho\hat
K\right)^{-1}\qquad \nn S&=&\left (I-\hat Re\, B\hat K+i\hat\rho\hat K\right)\left (I-\hat
Re\,B\hat K-i\hat\rho\hat K\right)^{-1}
 \eea

This approach provides a correct continuation of the amplitude below thresholds.

\subsection{The \boldmath$D$-matrix approach}

As we discussed above, the K-matrix approach can be considered as an effective way to calculate an
infinite sum of rescattering diagrams from the spectral integral equation.  The rescattering
diagrams can be divided into $K$-matrix blocks which describe a transition from one channel into
another one. Thus the rank of the $K$-matrix is defined by the number of the channels taken into
account explicitly. The key issue of the $K$-matrix approach is a factorization of vertices and
loop diagrams. The factorization is automatically fulfilled for the imaginary part, and in many
cases a contribution from the real part is neglected. When the vertices have a non-trivial energy
dependence, the real part can not be separated from the $K$-matrix block and another approach
should be used to calculate the amplitude. The most straightforward idea is to extract blocks which
describe a transition from one ``bare'' state to another one. Then, factorization is automatically
fulfilled for the pole terms.

Let us introduce the block $D_{\alpha\beta}$ which describes a transition between the bare state
$\alpha$ (but without the propagator of this state) and the bare state $\beta$ (with the propagator
of this state included). For such a block one can write the following equation:
 \bea
 D_{\alpha\beta}&=&
D_{\alpha\gamma}\sum\limits_j B^j_{\gamma\delta}d_{\delta\beta}+d_{\alpha\beta}
 \eea
 Or, in the matrix form,
 \be
 \hat D= \hat D\hat B\hat d+\hat d \qquad \hat D= \hat
d(I-\hat B\hat d)^{-1}
 \ee
 Here, the $\hat d$ is a diagonal matrix of the propagators
 \bea
 \hat d=diag\left
(\frac{1}{M^2_1-s},\frac{1}{M^2_2-s},\ldots,\frac{1}{M^2_N-s}
\right)\nn
 \eea
where $N$ is the number of resonant terms. The elements of the $\hat B$-matrix are equal to
 \bea
 \hat B_{\alpha\beta}&=&\sum\limits_j
 B^j_{\alpha\beta}= \quad \hfill \nn
 \sum\limits_j&&\hspace{-4mm }\int\limits^\infty_{(m_{1j}+m_{2j})^2}
\frac{ds'}{\pi}\frac{g^{R(\alpha)}_j\rho_j(s',m_{1j},m_{2j})g^{L(\beta)}_j}{s'-s-i0} \;.
 \eea
The $g^{R(\alpha)}_j$ and $g^{L(\alpha)}_j$ are right and left vertices for a transition from the
bare state $\alpha$ to the channel $j$. The function $B^j_{ab}$ depends on initial, intermediate
and final states and allows us to introduce for every transition a specific energy dependence and
regularization procedure.

For the resonance transition the right and left vertices are the same:
\be
g^{R(\alpha)}_j=g^{L(\alpha)}_j=g^{(\alpha)}_j \ee

The scattering amplitude between channels $i$ and $j$ which are taken into
account in the rescattering has the form \be A_{ij}=g_i^{(\alpha)}\hat
D_{\beta\gamma}g_j^{(\gamma)}. \ee

\begin{figure*}[pt]
\begin{center}
\begin{tabular}{cccc}
\hspace{-1mm}\includegraphics[width=0.325\textwidth,height=0.22\textheight]{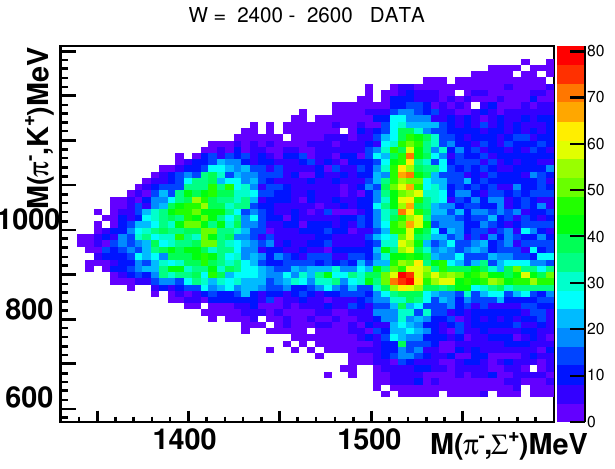}&
\hspace{-3mm}\includegraphics[width=0.325\textwidth,height=0.22\textheight]{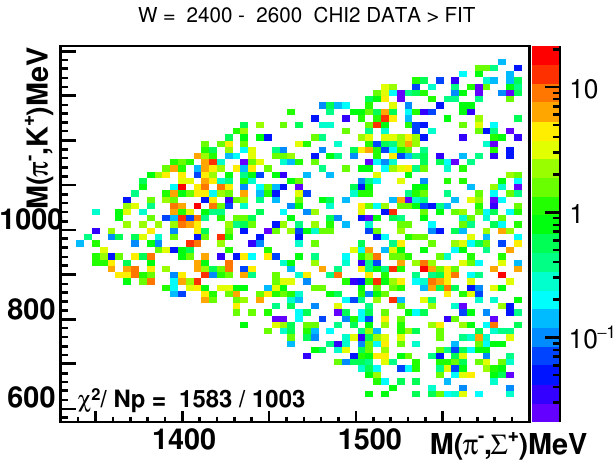}&
\hspace{-3mm}\includegraphics[width=0.325\textwidth,height=0.22\textheight]{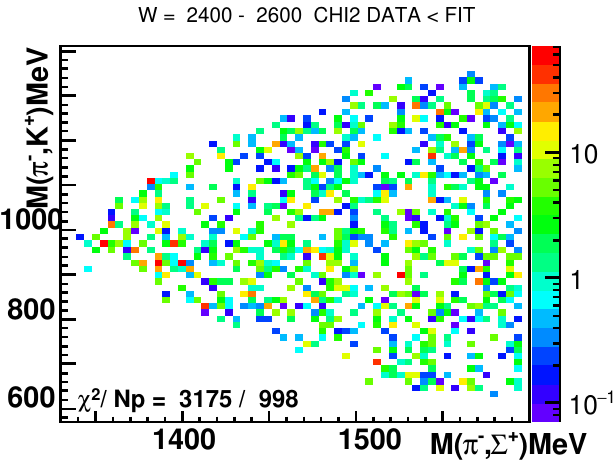}
\end{tabular}
\end{center}
\caption{\label{diff1} $\gamma p \to K^+ \pi^- \Sigma^+$ for $2400<W=M_{\gamma p}<2600$,MeV: $M(\pi^-K^+)$ versus $M(\pi^-\Sigma^+)$  two-dimensional mass distributions, upper row: reconstructed data without acceptance correction, middle/lower row: $\chi^2$-distributions for the case where the data exceed the fit (middle row) and where the fit exceeds the data (lower row). 
 \vspace{-2mm}}
\begin{center}
\begin{tabular}{cccc}
\hspace{-1mm}\includegraphics[width=0.325\textwidth,height=0.22\textheight]{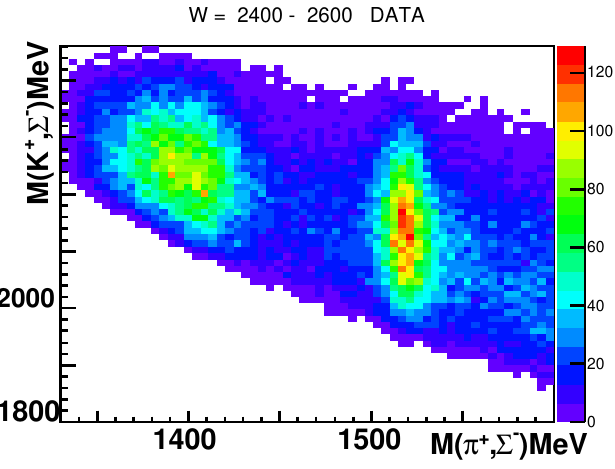}&
\hspace{-3mm}\includegraphics[width=0.325\textwidth,height=0.22\textheight]{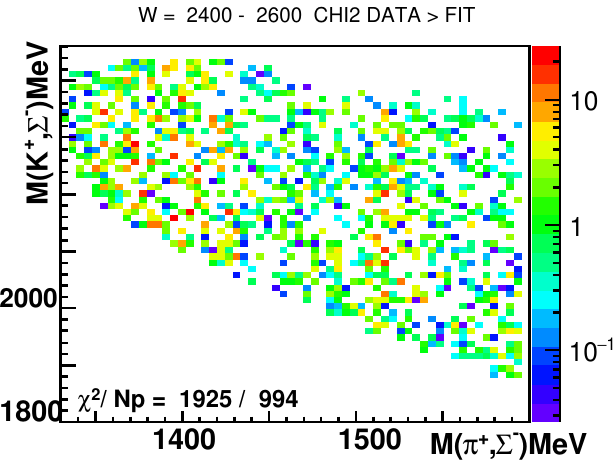}&
\hspace{-3mm}\includegraphics[width=0.325\textwidth,height=0.22\textheight]{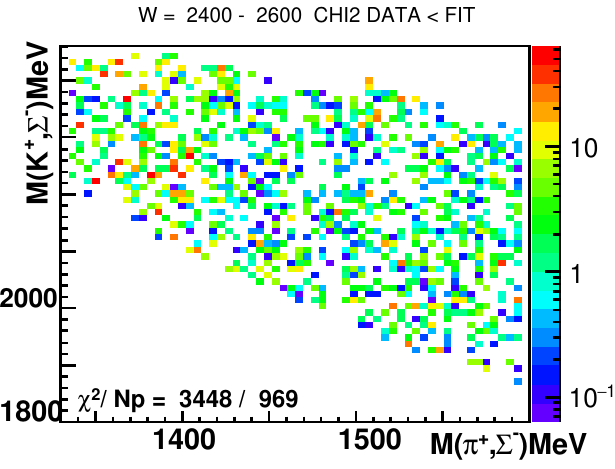}
\end{tabular}
\end{center}
\caption{\label{diff2} 
$\gamma p \to K^+ \pi^+ \Sigma^-$  for $2400<W=M_{\gamma p}<2600$,MeV: $M(K^+\Sigma^-)$ versus $M(\pi^+\Sigma^-)$   two-dimensional mass distributions, upper row: reconstructed data without acceptance correction, middle/lower row: $\chi^2$-distributions for the case where the data exceed the fit (middle row) and where the fit exceeds the data (lower row).
\vspace{-2mm}}
\end{figure*}
\begin{figure}[pt]
 \bc
\begin{tabular}{ccc}
\hspace{-1mm}\includegraphics[width=0.24\textwidth]{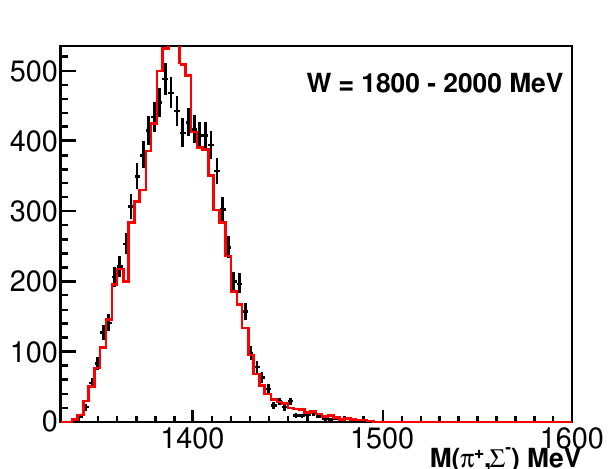}&
\hspace{-2mm}\includegraphics[width=0.24\textwidth]{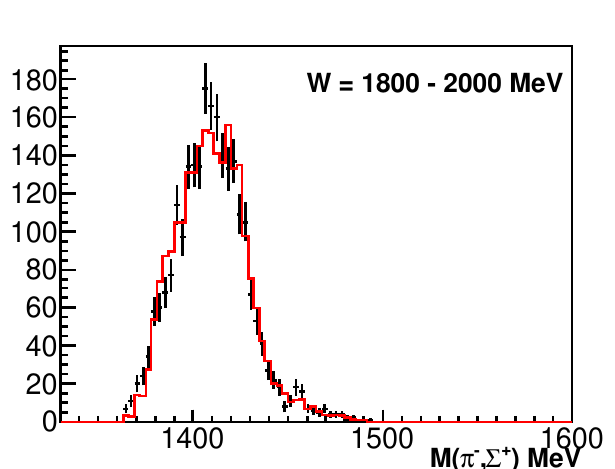}\\
\hspace{-1mm}\includegraphics[width=0.24\textwidth]{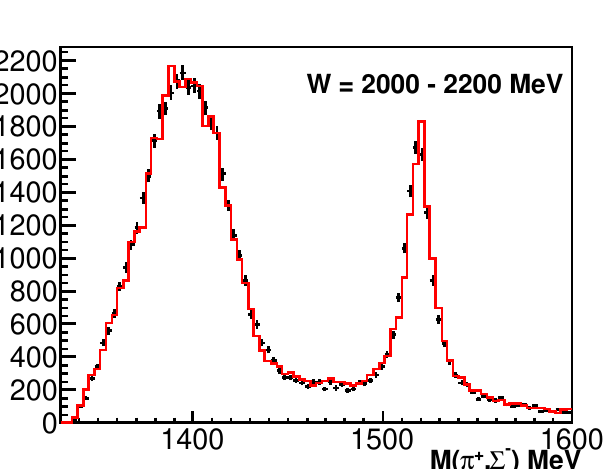}&
\hspace{-2mm}\includegraphics[width=0.24\textwidth]{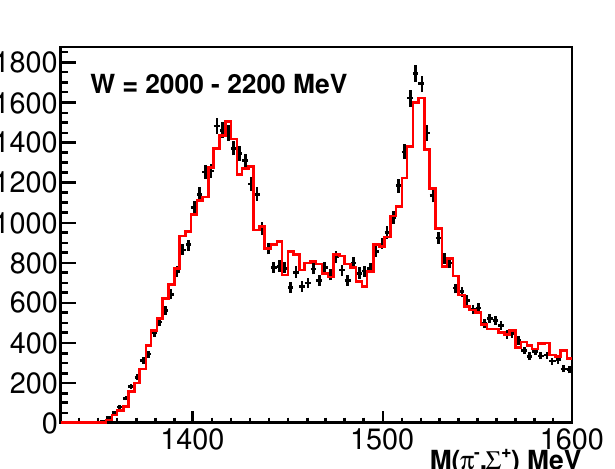}\\
\hspace{-1mm}\includegraphics[width=0.24\textwidth]{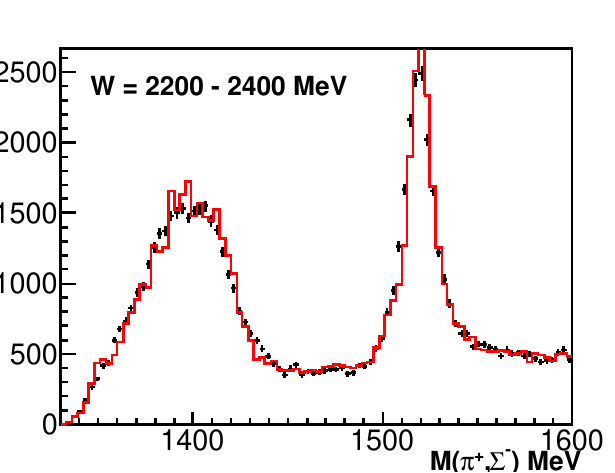}&
\hspace{-2mm}\includegraphics[width=0.24\textwidth]{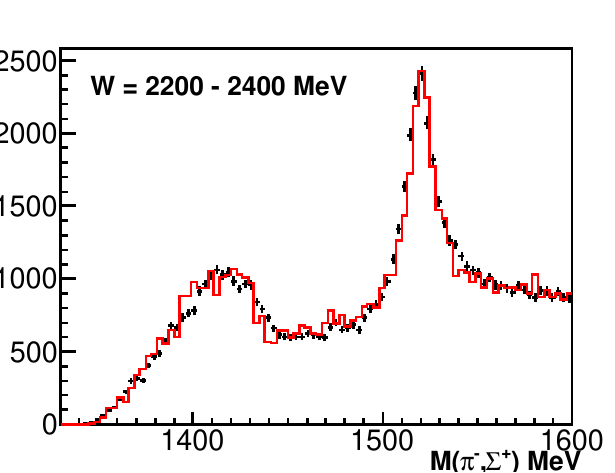}\\
\hspace{-1mm}\includegraphics[width=0.24\textwidth]{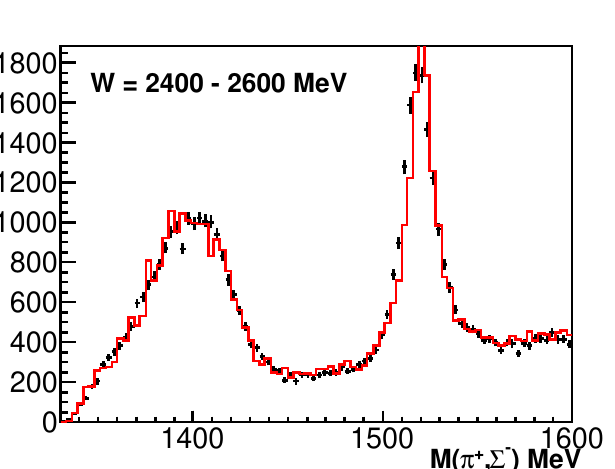}&
\hspace{-2mm}\includegraphics[width=0.24\textwidth]{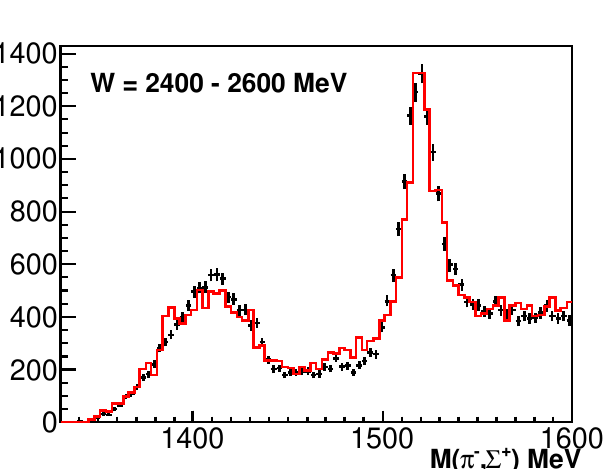}\\
\hspace{-1mm}\includegraphics[width=0.24\textwidth]{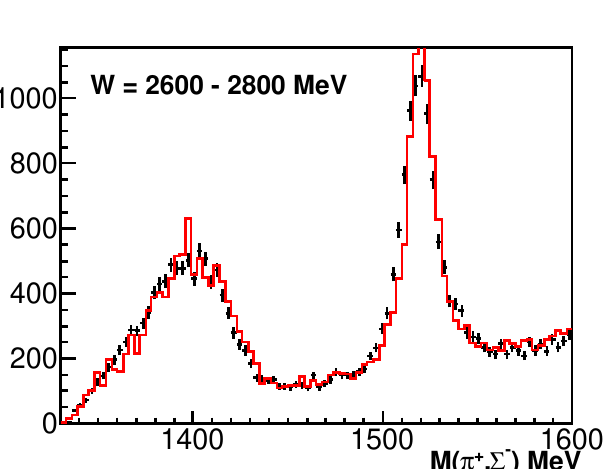}&
\hspace{-2mm}\includegraphics[width=0.24\textwidth]{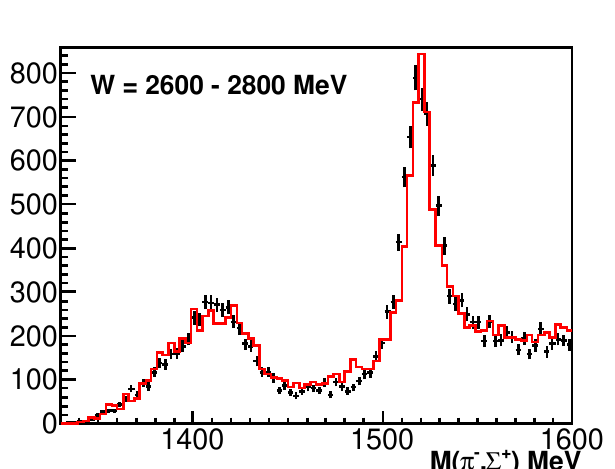}\\
\end{tabular} \vspace{-3mm}\ec
\caption{\label{CLAS}The $\pi^+\Sigma^-$ (left), $\pi^-\Sigma^+$ (right) invariant mass
distributions from the reaction $\gamma p\to  K^+\Sigma\pi$~\cite{Moriya:2013eb}, given as number
of events per 30\,MeV. The data are fitted in a likelihood fit to individual events. The fit,
represented by the histograms (red), uses one pole to describe $\Lambda(1405)$.
  } \vspace{-2mm}
\end{figure}
\begin{figure}[pt]
 \bc
\begin{tabular}{cc}
\hspace{-2mm}\includegraphics[width=0.25\textwidth]{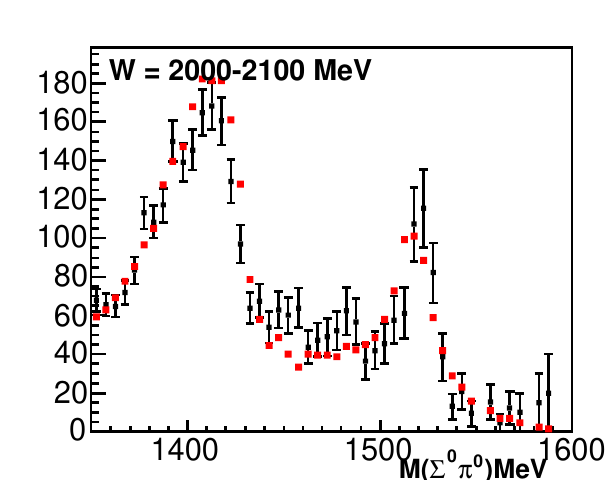} &
\hspace{-4mm}\includegraphics[width=0.25\textwidth]{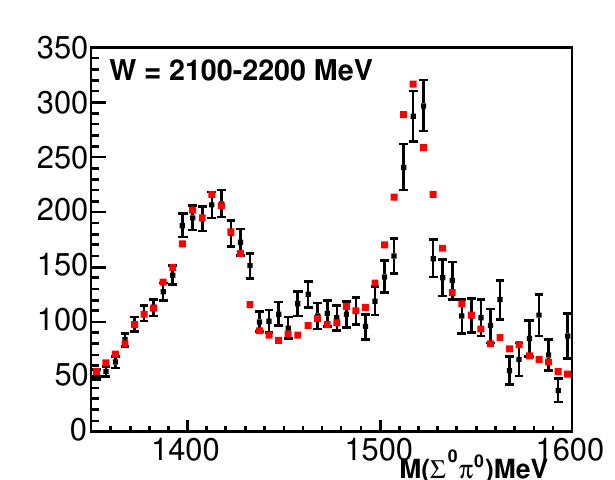}\\
\hspace{-2mm}\includegraphics[width=0.25\textwidth]{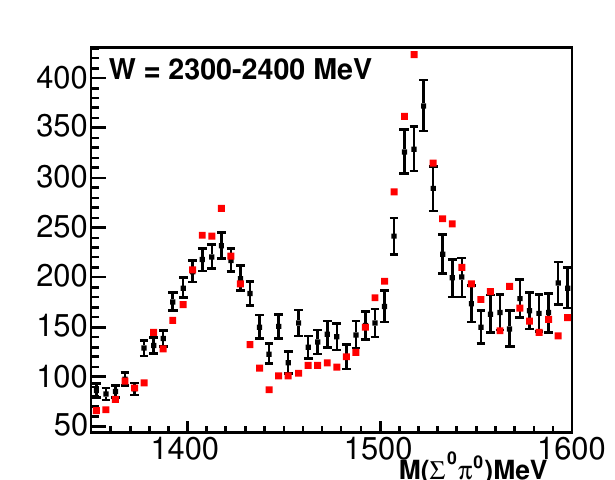}&
\hspace{-4mm}\includegraphics[width=0.25\textwidth]{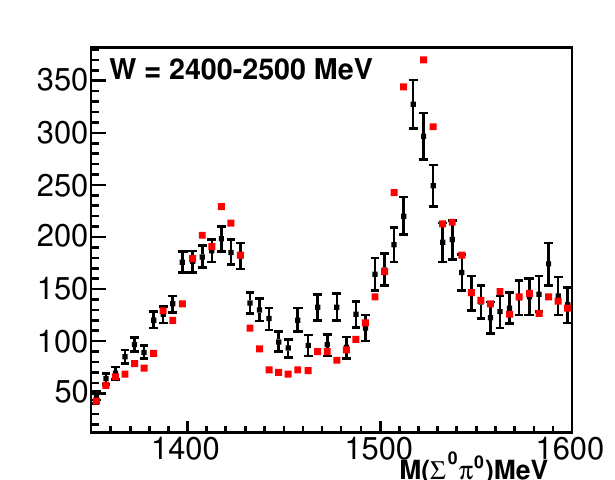} \\ 
\hspace{-2mm}\includegraphics[width=0.25\textwidth]{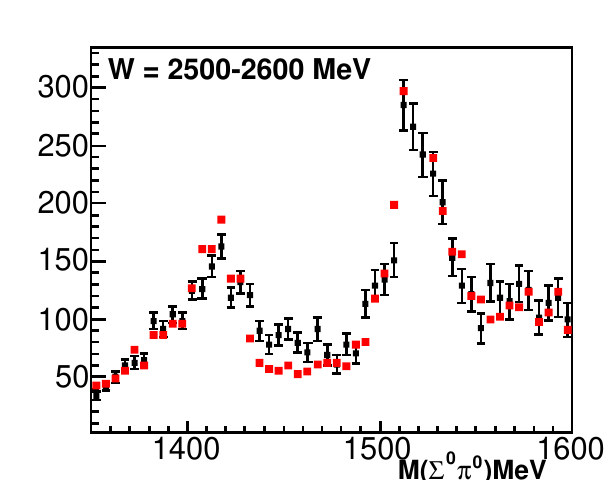}&
\hspace{-4mm}\includegraphics[width=0.25\textwidth]{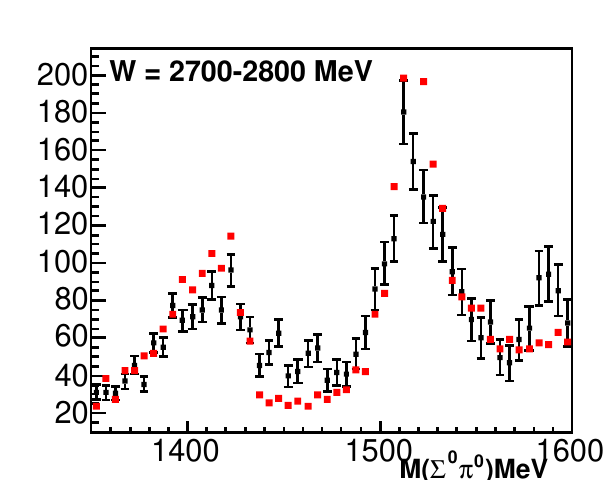}\\
\end{tabular} \vspace{-3mm}\ec
\caption{\label{CLAS2} The $\pi^0\Sigma^0$ mass distribution~\cite{Moriya:2013eb} is given as
weighted number of events per 5\,MeV. The data are not included in the fit, the prediction is
represented by (red) dots.
  } \vspace{-2mm}
\end{figure}

\subsection{\boldmath$\Lambda(\frac12^-)$ and $\Sigma(\frac12^-)$ partial waves parameterizations}

We are interested in the amplitude behavior in the region from the $\pi\Sigma$ threshold to
$\sqrt{s} \sim 1.5$\,GeV. Hence both $I=0$ and $I=1$ amplitudes could contain one or two poles. The
fit should tell us where the poles of the amplitudes are located. The $\Lambda(\frac12^-)$
amplitude is described by a five-channel amplitude with possible decays to $\pi^-\Sigma^+$,
$\pi^0\Sigma^0$, $\pi^+\Sigma^-$, $K^-p$ and $K^0n$. The constructed amplitudes take into account
isotopic mass differences (threshold positions) but neglect Coulomb interactions. It means that a
one-pole D-Matrix amplitude depends on two real coupling constants ($g_{\pi\Sigma}$ and $g_{KN}$)
and a bare mass value $M$. These parameters are defined in the fit. The two-pole amplitude depends
on six fitting parameters. The $\Sigma(\frac12^-)$ amplitude has additionally the $\Lambda \pi^0$
channel, so we have a six-channel amplitude. We use a two-pole parametrization for this amplitude
which depends on eight fit parameters.

\section{\label{Fits}Fits to the data}

The mass of $\Lambda(1405)$ falls below the $K^-p$ threshold. In $K^-p$ induced reactions only the
high-mass part of $\Lambda(1405)$ can be produced. An important role for the study of
$\Lambda(1405)$ is hence provided by the CLAS results on $\gamma p\to K^+\Sigma^+\pi^-$,
$K^+\Sigma^0\pi^0$, and $K^+\Sigma^-\pi^+$~\cite{Moriya:2013eb} where the full $\Lambda(1405)$
shape can be studied. Fig.~\ref{diff1} (left) and \ref{diff2} (left) show selected two-dimensional mass distributions:
$M_{\pi^-K^+}$ versus $M_{\pi^-\Sigma^+}$ and $M_{K^+\Sigma^-}$ versus $M_{\pi^+\Sigma^-}$ for a $\gamma p$ invariant mass in the 2400 - 2600\,MeV range.  In both figures,
a vertical band is seen at $M_{\Sigma^+\pi^-}$\,GeV or
$M_{\Sigma^-\pi^+}=1.52$\,GeV: the $\Lambda(1520)$. At low masses, a broad enhancement due to 
$\Sigma(1385)$ and $\Lambda(1405)$ is seen which both decay into $\Sigma^\pm\pi^\mp$. 
A horizonzal band in Fig.~\ref{diff1} evidences $K^*$ production. The resonances
$K^*$, $\Sigma(1385)$ and $\Lambda(1520)$  are described by
relativistic Breit-Wigner amplitudes with masses and widths compatible with the PDG central 
values~\cite{Tanabashi:2018oca}. The $K^*$ band interferes with
$\Sigma(1385)$, $\Lambda(1405)$, and $\Lambda(1520)$.

In Fig.~\ref{diff2} (left) , the $M_{K^+\Sigma^-}$ invariant mass is plotted against $M_{\Sigma^-\pi^+}$.
There are no longer striking horizontal bands which would indicate $\Sigma^+K^+$ resonances. There is also
no $K^+\pi^+$ band which would show up as a band in the counterdiagonal. 

The data were fitted event by event in a likelihood fit. The center and right subfigures in
Figs.~\ref{diff1} and \ref{diff2} show the $\chi^2$ per bin for events in which the data exceed the
fit and for events in which the fit exceeds the data. The $\chi^2$ of the fit is moderate: it is
41320 for 16076 cells. However, no significant pattern is seen in the difference plots. Hence we
believe the fit to be acceptable.

Figure~\ref{CLAS} shows the two $\Sigma^\pm\pi^\mp$ mass distributions and the BnGa fit. The
$\Lambda(1520)$ resonance is clearly seen. The low-mass structure contains contributions from
$\Lambda(1405)$ and from $\Sigma(1385)$. 
The result of the fit was then used to predict the
$\Sigma^0\pi^0$ mass distribution for events from $\gamma p\to K^+\Sigma^0\pi^0$.  Data and
prediction are shown in Fig.~\ref{CLAS2}. The fit identifies the two components reliably; the
prediction for the $\pi^0\Sigma^0$ mass distribution is very good: this distribution contains no
$\Sigma(1385)$ since the decay $\Sigma(1385)\to \pi^0\Sigma^0$ is forbidden.

\begin{figure}[b]
\vspace{-2mm}
\bc
\includegraphics[width=0.45\textwidth,height=0.35\textwidth]{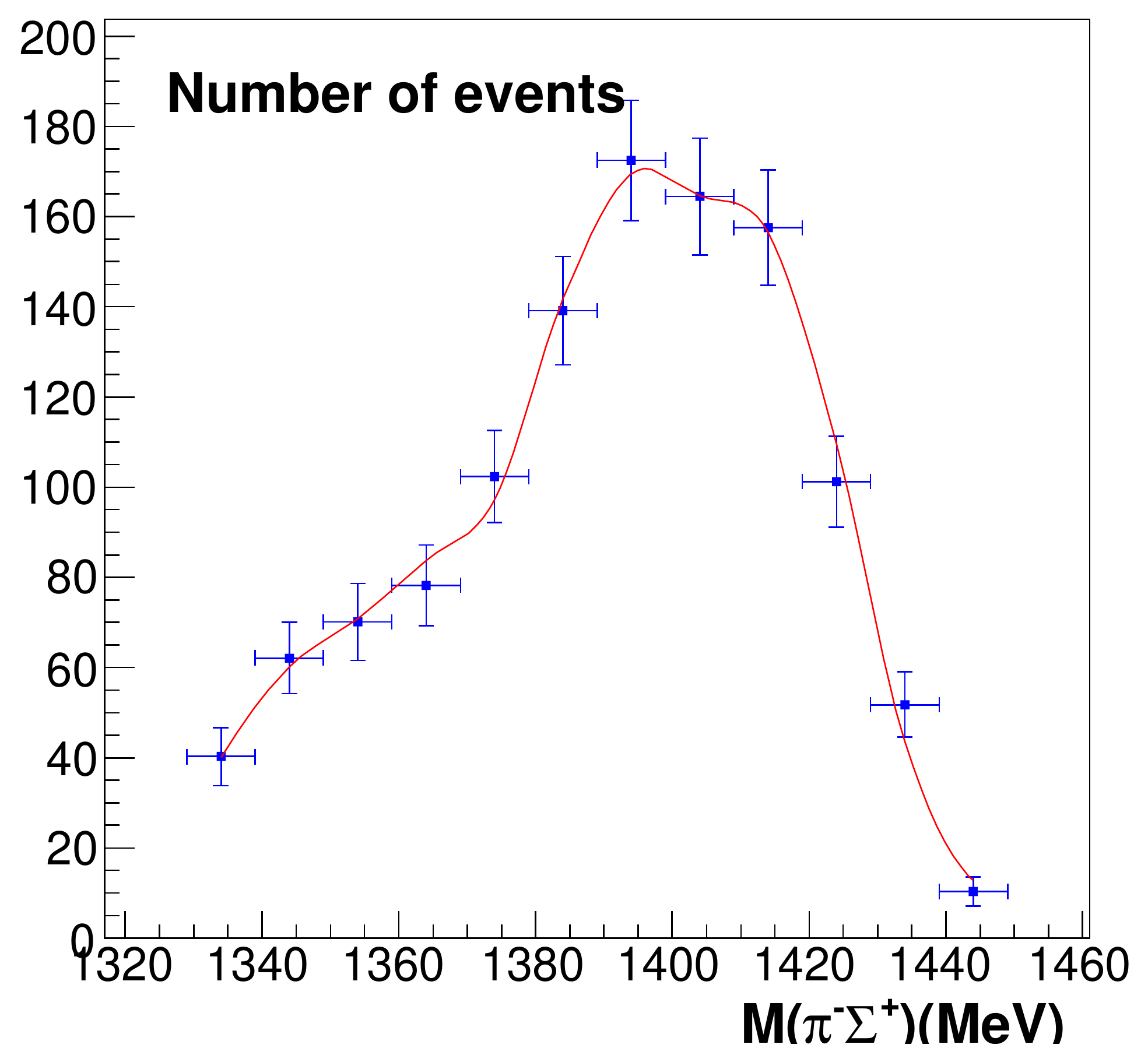}\vspace{-2mm}
\ec
\caption{\label{Hemingway}$\Sigma^+\pi^-$ mass projection from the reaction
$K^-p\to\pi^-\pi^+\pi^{\mp}\Sigma^{\pm}$ for events with $M_{\pi^+\pi^{\pm}\Sigma^{\mp}}$
compatible with $\Sigma(1670)3/2^-$~\cite{Hemingway:1984pz}. Shown is the number of events
per 10\,MeV.  }
\end{figure}

Before the CLAS data became available, the full $\Lambda(1405)$ mass distribution was accessible from
old bubble chamber data on $K^-p\to\pi^-\pi^+\pi^{\pm}\Sigma^{\mp}$~\cite{Hemingway:1984pz}. The
$\Lambda(1405)$ was observed in the $K^-p\to\pi^-\Sigma^+(1670)3/2^-$, $\Sigma^+(1670)3/2^- \to$
$\pi^+\Lambda(1405)$ cascade, with $\Lambda(1405)1/2^-\to\pi^\pm\,\Sigma^{\mp}$. In the fit, a
$\approx25$\% fraction of $\Sigma^0(1385)$ was admitted. The data are well reproduced by our
fit with $\chi^2/(N_{\rm data}-N_{\rm param.})= 3.3/(12-7)$ (see Fig.~\ref{Hemingway}).

\begin{figure*}[pt]
\begin{center}
\begin{tabular}{cccc}
\hspace{-2mm}\includegraphics[width=0.25\textwidth]{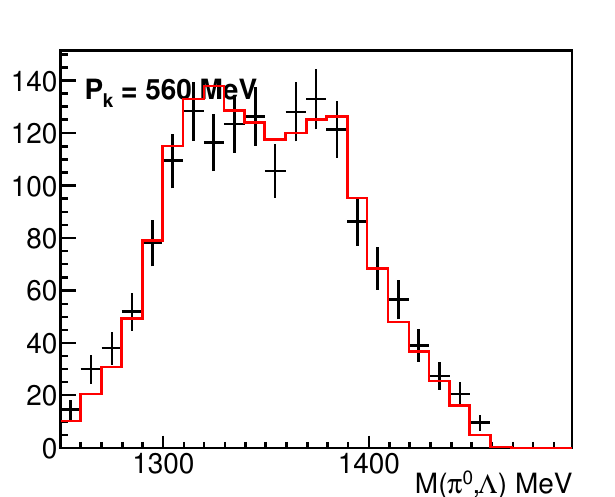}&
\hspace{-4mm}\includegraphics[width=0.25\textwidth]{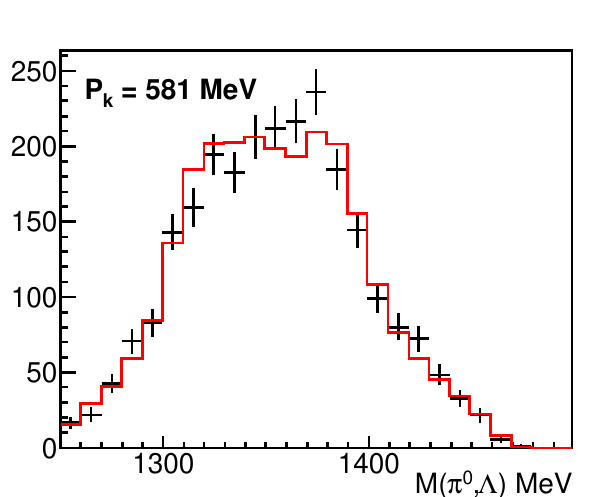}&
\hspace{-1mm}\includegraphics[width=0.25\textwidth]{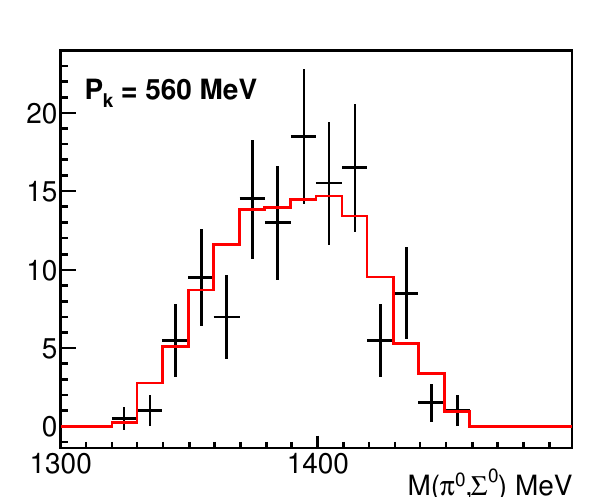}&
\hspace{-4mm}\includegraphics[width=0.25\textwidth]{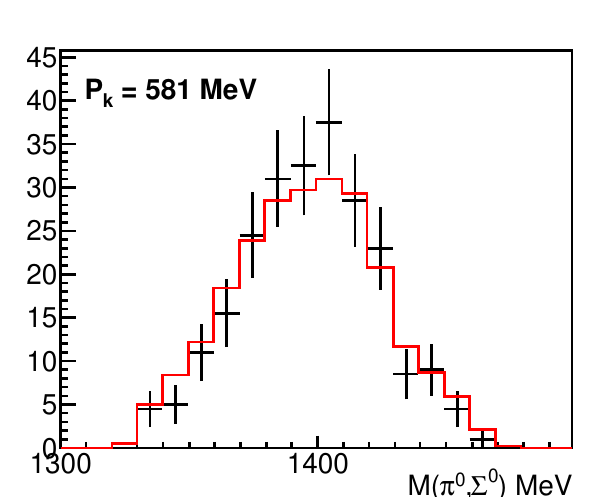}\\[-1ex]
\hspace{-2mm}\includegraphics[width=0.25\textwidth]{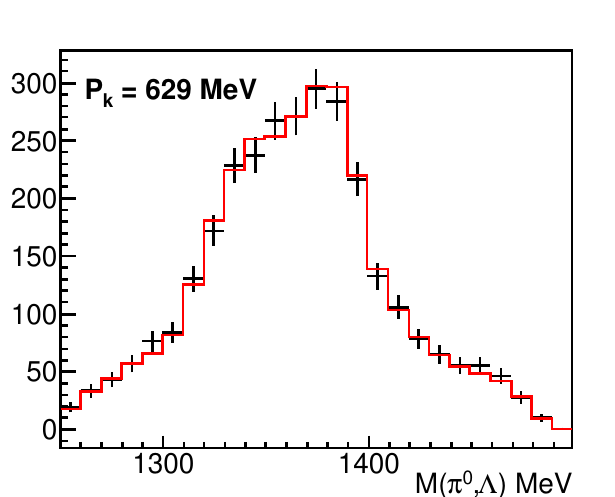}&
\hspace{-4mm}\includegraphics[width=0.25\textwidth]{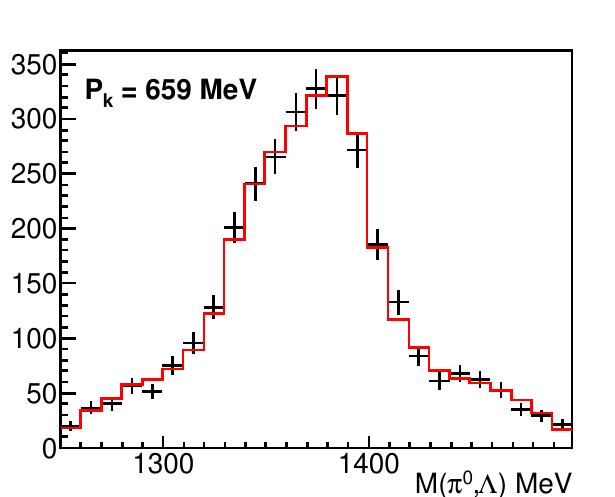}&
\hspace{-1mm}\includegraphics[width=0.25\textwidth]{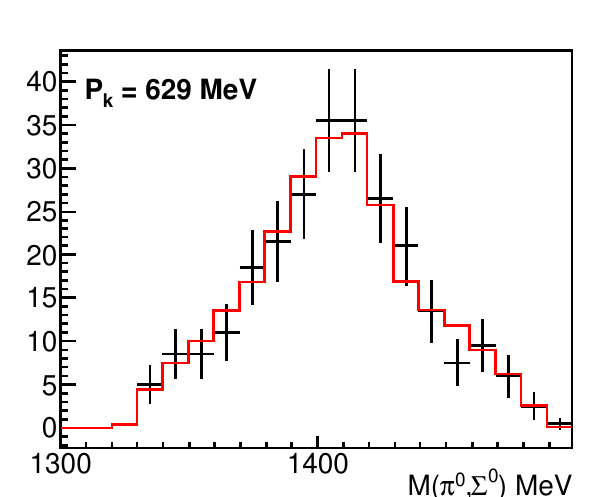}&
\hspace{-4mm}\includegraphics[width=0.25\textwidth]{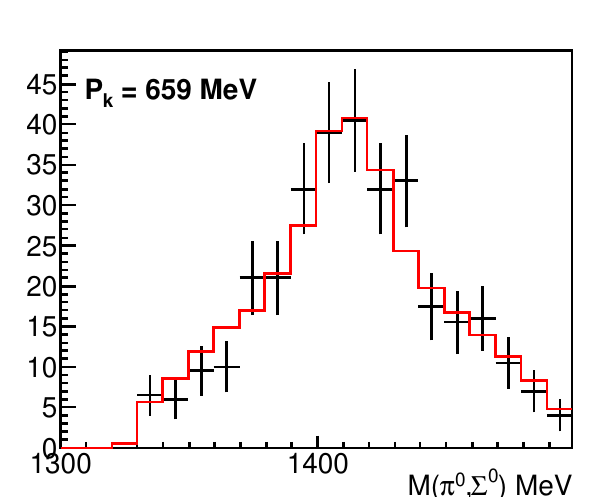}\\[-1ex]
\hspace{-2mm}\includegraphics[width=0.25\textwidth]{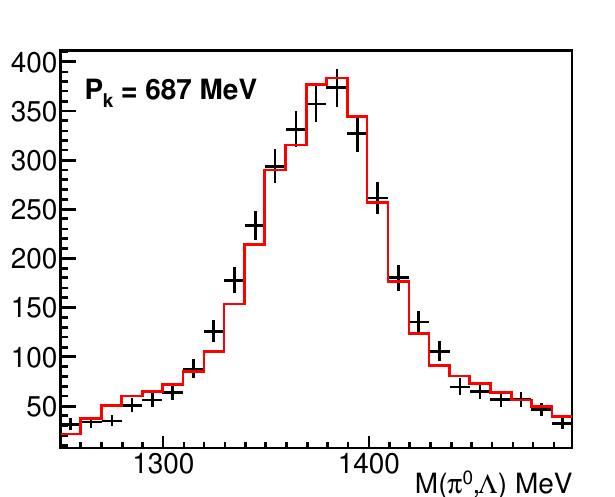}&
\hspace{-4mm}\includegraphics[width=0.25\textwidth]{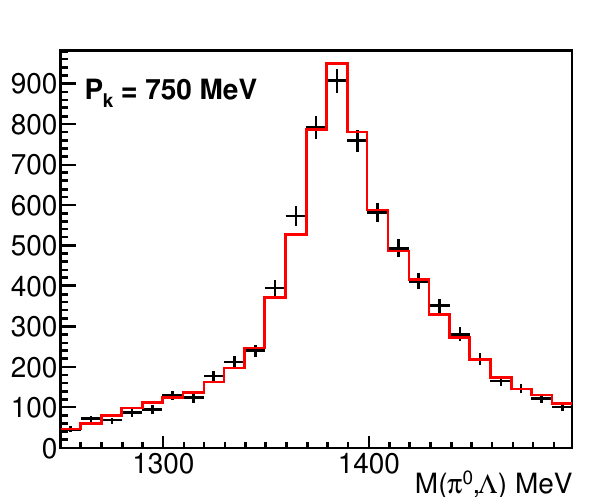}&
\hspace{-1mm}\includegraphics[width=0.25\textwidth]{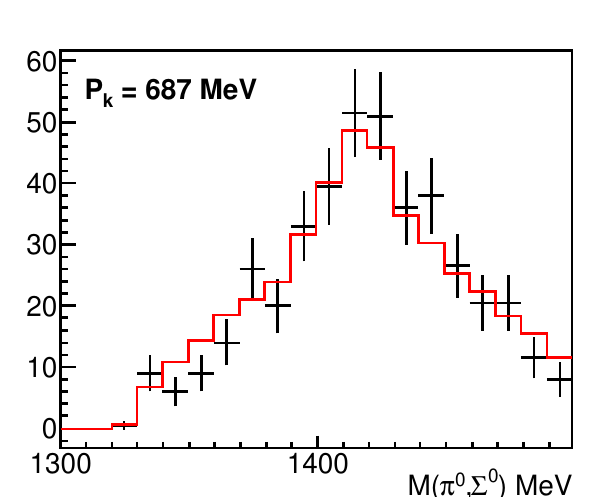}&
\hspace{-4mm}\includegraphics[width=0.25\textwidth]{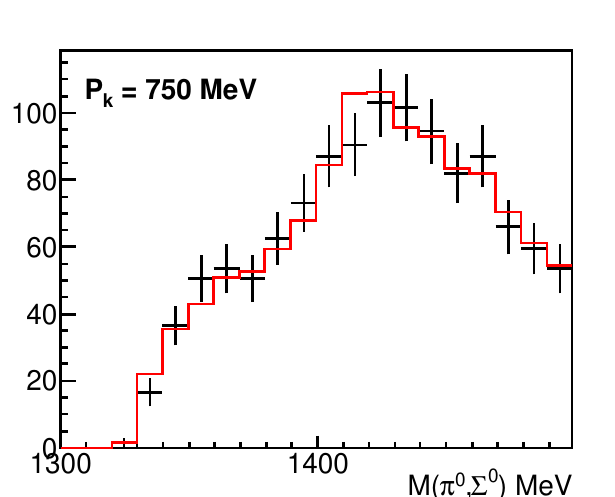}\\[-3ex]
\end{tabular} \end{center} \caption{\label{BNL} The Crystal Ball
data on $K^-p\to\pi^0\pi^0\Lambda$ (left) and $K^-p\to\pi^0\pi^0\Sigma^0$
(right)~\cite{Prakhov:2004an} shown as black data points. One pole solution
is presented as histogram (red).\vspace{-2mm}}
\end{figure*}

The Crystal Ball Collaboration at BNL studied the reactions $K^-p\to\pi^0\pi^0\Lambda$ and
$K^-p\to\pi^0\pi^0\Sigma^0$~\cite{Prakhov:2004an}. The events were fitted maximizing the likelihood
in an event-by-event fit. Figure~\ref{BNL} shows the $\pi^0\Lambda$ and $\pi^0\Sigma^0$ invariant
mass distributions and the fit. In the $\pi^0\Lambda$ distribution, the $\Sigma(1385)$ dominates
the reaction, a peak in the $\pi^0\Sigma^0$ mass distribution provides evidence for
$\Lambda(1405)$. The data are well reproduced by the fit.

\begin{figure}[pt]
\begin{center}
\begin{tabular}{ccc}
\includegraphics[width=0.22\textwidth,height=0.17\textwidth]{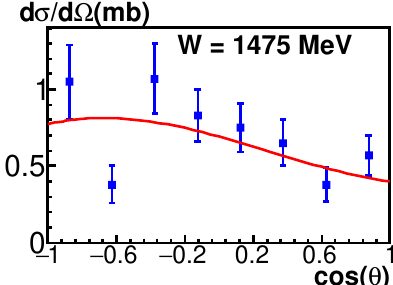}&
\includegraphics[width=0.22\textwidth,height=0.17\textwidth]{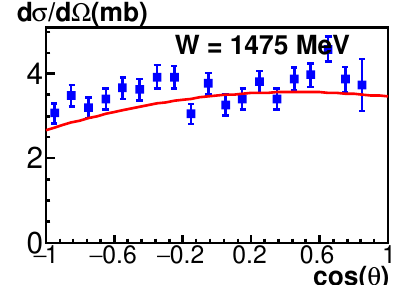}\\[+1ex]
\includegraphics[width=0.22\textwidth,height=0.17\textwidth]{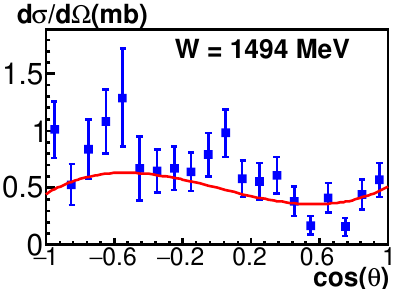}&
\includegraphics[width=0.22\textwidth,height=0.17\textwidth]{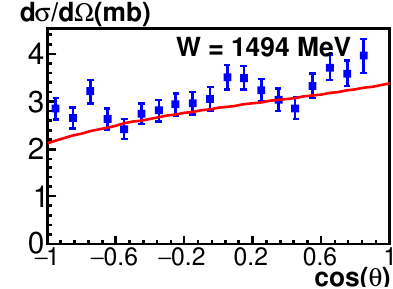}\\[+1ex]
\includegraphics[width=0.22\textwidth,height=0.17\textwidth]{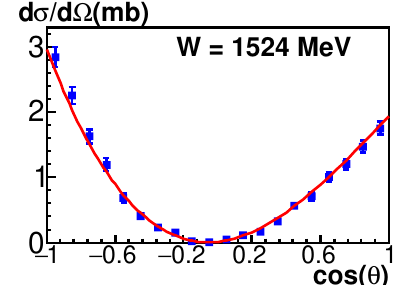}&
\includegraphics[width=0.22\textwidth,height=0.17\textwidth]{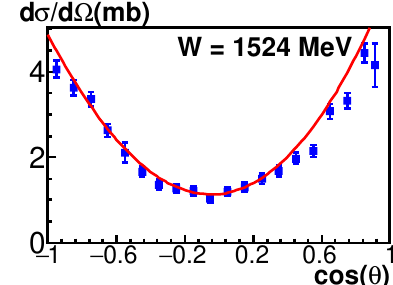}\\[-2.5ex]
\end{tabular}
\end{center}
\caption{\label{diff} Selected differential cross sections for $K^-p\to K^-p$ (left) and
$K^-p\to\bar K^0n$ (right) from \cite{Mast:1975pv}. Data were taken between 1464 and 1548\,MeV invariant
mass and reported in 25 bins. Shown are three bins. The fit is given by the line. \vspace{-2mm}}
\end{figure}

$K^-p$ scattering starts at 1432\,MeV, above the nominal mass of $\Lambda(1405)$. Nevertheless,
kaon-induced reactions provide significant constraints on the $I(J^P)=0(1/2^-)$-amplitude.
Figure~\ref{diff} shows the differential cross section for $K^-p\to K^-p$ and $K^-p\to\bar K^0n$
from \cite{Mast:1975pv} in selected bins of the invariant mass. The data are reasonably well
described.

\begin{figure}[pt]
\begin{center}
\begin{tabular}{cc}
\hspace{-2mm}\includegraphics[width=0.25\textwidth,height=0.173\textwidth]{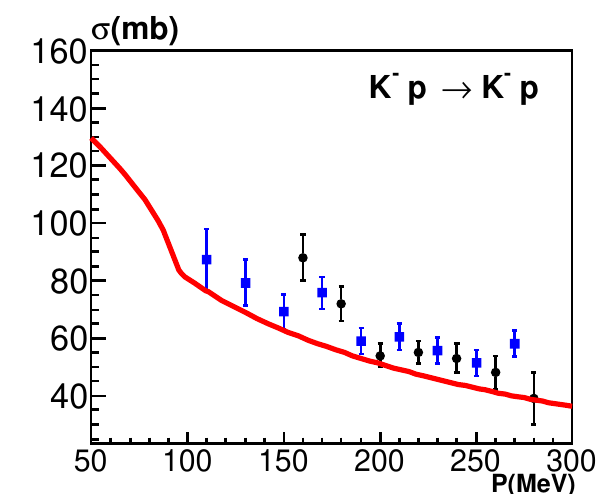}&
\hspace{-2mm}\includegraphics[width=0.25\textwidth,height=0.173\textwidth]{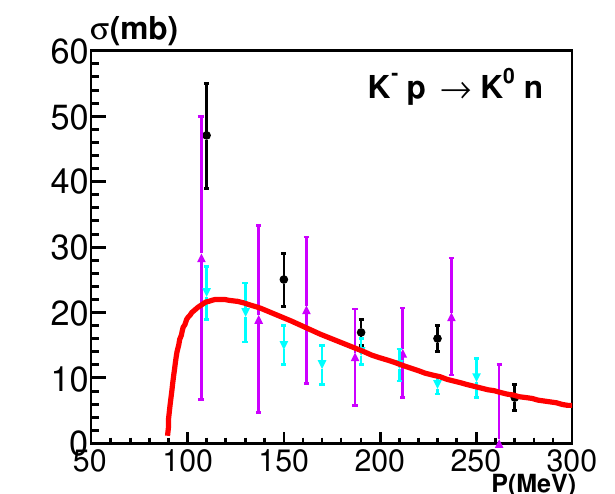}\\
\hspace{-2mm}\includegraphics[width=0.25\textwidth,height=0.173\textwidth]{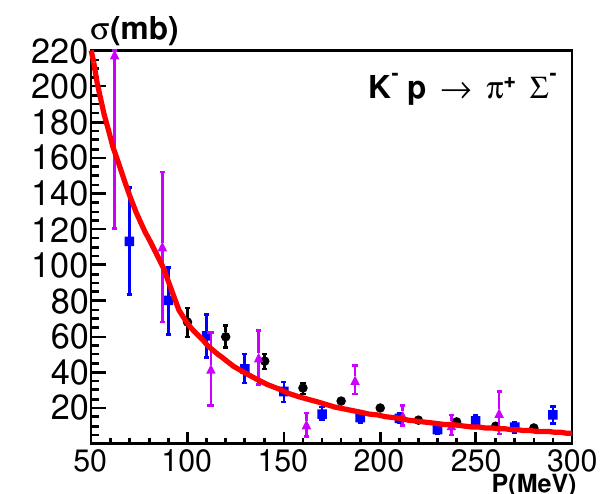}&
\hspace{-2mm}\includegraphics[width=0.25\textwidth,height=0.173\textwidth]{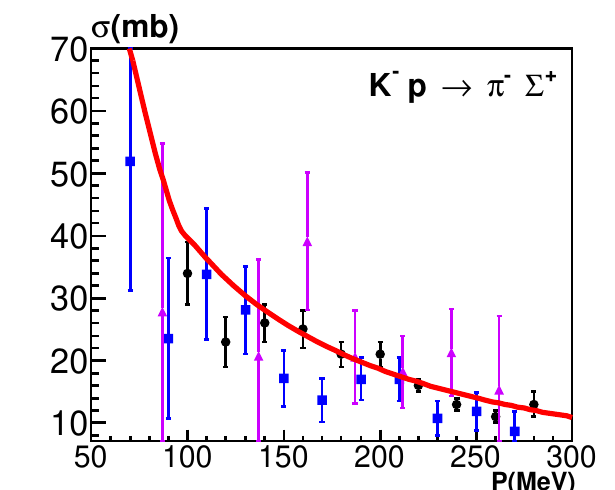}\\
\hspace{-2mm}\includegraphics[width=0.25\textwidth,height=0.173\textwidth]{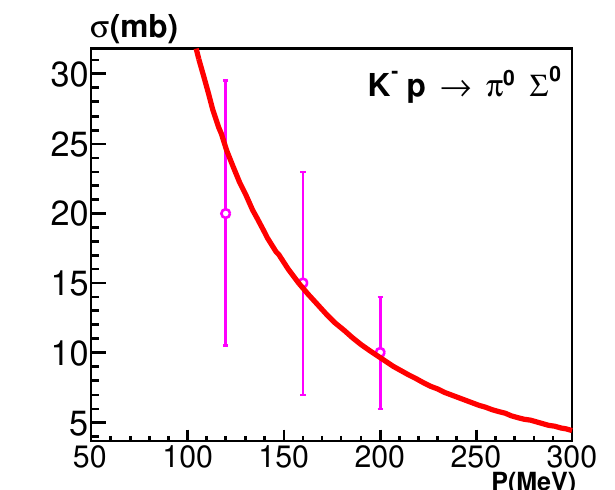}&
\hspace{-2mm}\includegraphics[width=0.25\textwidth,height=0.173\textwidth]{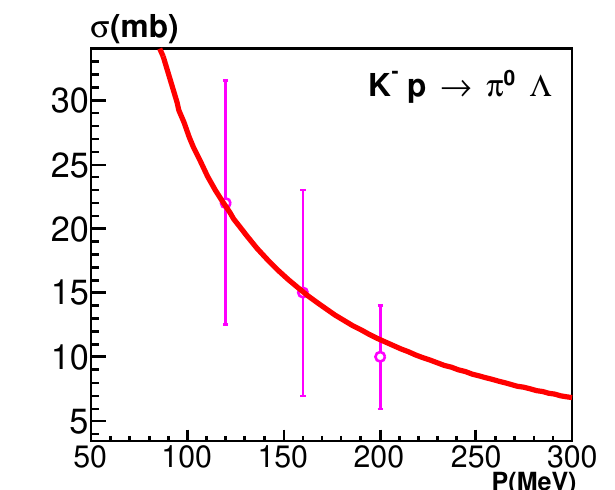}\\[-2ex]
\end{tabular}
\end{center}
\caption{\label{TCS} The total cross sections for $K^-p$ induced reactions: $K^-p\to
K^-p$, $K^-p\to\bar K^0n$, $K^-p\to\pi^0\Lambda$, $K^-p\to\pi^+\Sigma^-$, $K^-p\to\pi^0\Sigma^0$,
$K^-p\to\pi^-\Sigma^+$~\cite{Humphrey:1962zz,Watson:1963zz,Sakitt:1965kh,Ciborowski:1982et}. The
single pole $\Lambda(1405)$ fit is given as the red curve.\vspace{-2mm}}
\end{figure}
\begin{figure}[pt]
\begin{center}
\begin{tabular}{cc}
& \\[-4ex]
\hspace{-2mm}\includegraphics[width=0.25\textwidth]{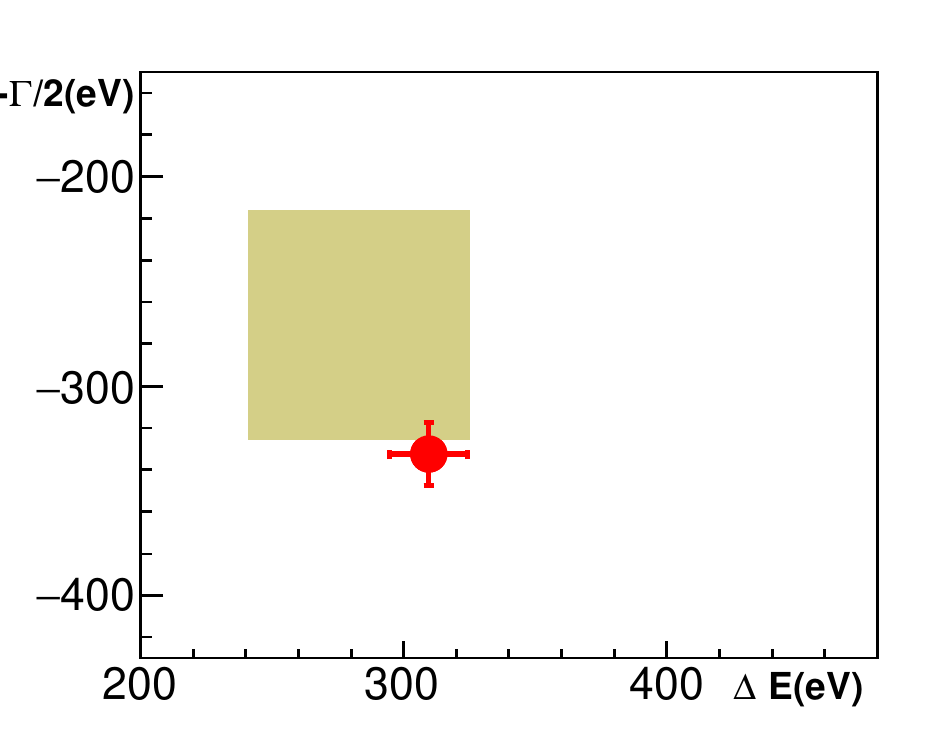}&
\hspace{-4mm}\includegraphics[width=0.25\textwidth]{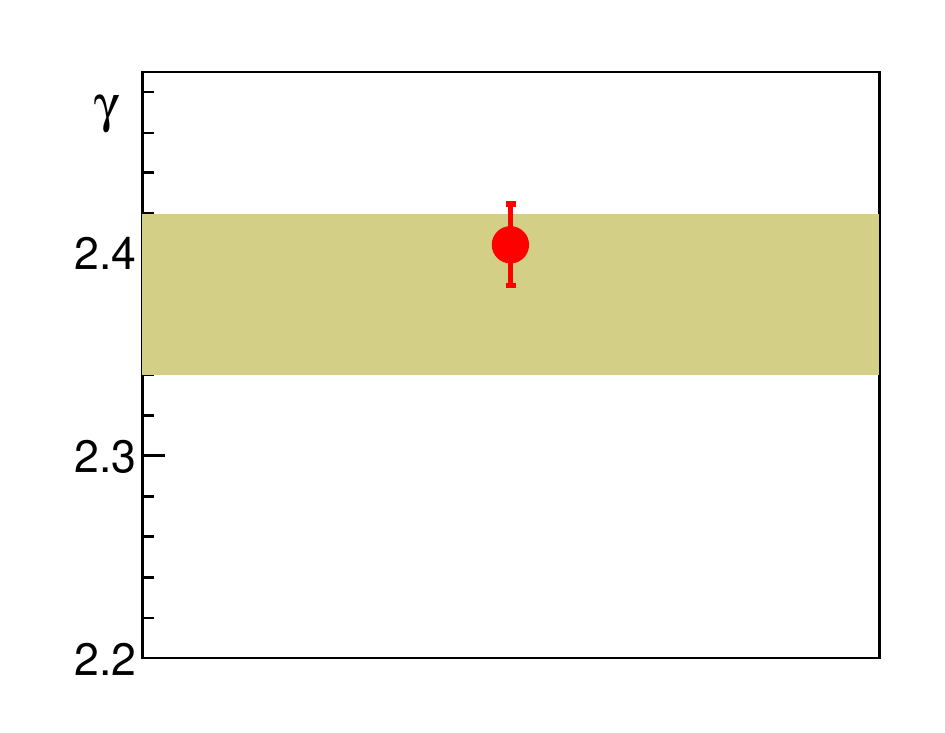}\\[-1ex]
\hspace{-2mm}\includegraphics[width=0.25\textwidth]{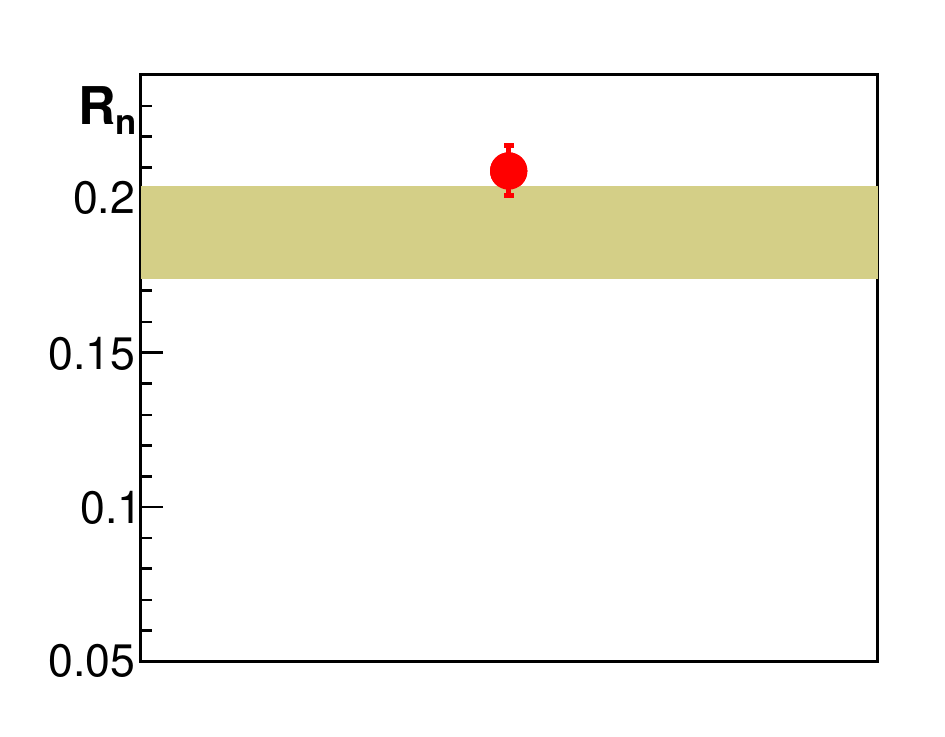}&
\hspace{-4mm}\includegraphics[width=0.25\textwidth]{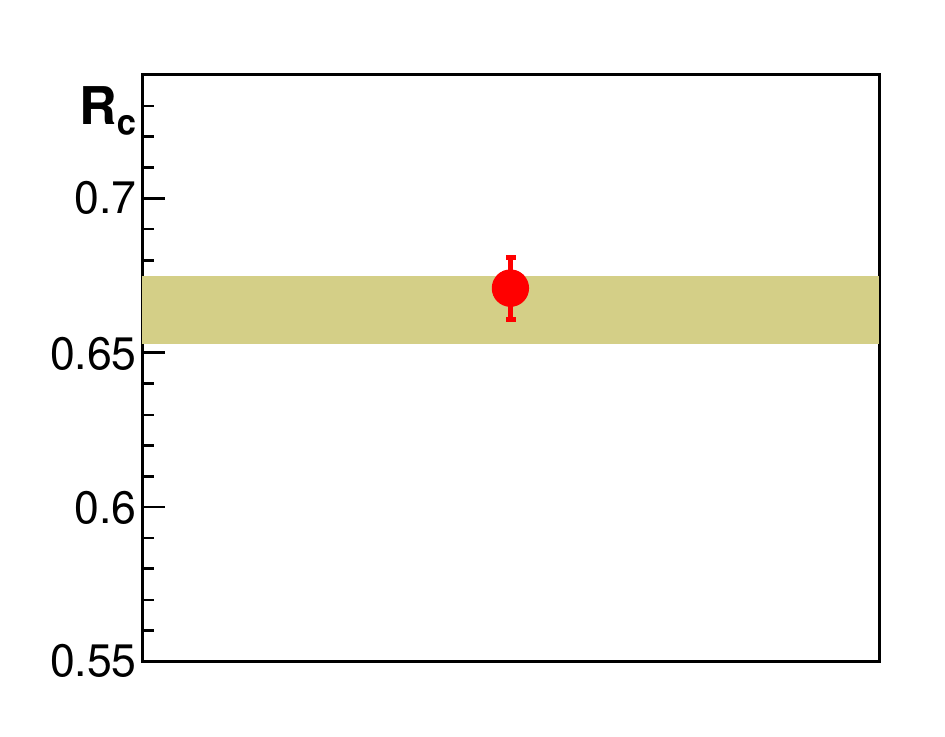}\\[-4ex]
\end{tabular}
\end{center}
\caption{\label{Atom} The threshold quantities defined in
Eqs.~(\ref{atom}) with errors (shaded areas). The single-pole $\Lambda(1405)$ fit is given in red.
The shaded area represents the fit uncertainty.}
\end{figure}
%
%
%
%
%

Figure~\ref{TCS} shows the total cross sections for $K^-p$ induced reactions: $K^-p\to K^-p$,
$K^-p\to\bar K^0n$, $K^-p\to\pi^0\Lambda$, $K^-p\to\pi^+\Sigma^-$, $K^-p\to\pi^0\Sigma^0$,
$K^-p\to\pi^-\Sigma^+$~\cite{Humphrey:1962zz,Watson:1963zz,Sakitt:1965kh,Ciborowski:1982et}. The
data are restricted to the low-mass region region, with $K^-$ laboratory momentum $P_{\rm
lab}<300$\,MeV, where the $P$-wave scattering amplitude can be neglected. Note that the fit curve
for the elastic scattering total cross section is rather determined by the differential cross
section of the data from \cite{Mast:1975pv} and hardly influenced by the data on the total cross
section.

The fits are constrained by properties of the $K^-p$ system at rest. The SIDDHARTA experiment at
DA$\Phi$NE determined the energy shift and width of the 1S level of the kaonic hydrogen
atom~\cite{Bazzi:2011zj,Bazzi:2012eq}. The values (Eq.~\ref{a}) are related to the $K^-p$
scattering length via the modified Deser-type formula \cite{Meissner:2004jr}:
\begin{align}
 \Delta E -i\Gamma/2=-2\alpha^3\mu^2_ca_{K^-p}
 \left[1-2a_{K^-p}\alpha\mu_c(\ln \alpha -1)\right]\,,
\end{align}
where $\alpha \simeq 1/137$ is the fine-structure constant, $\mu_c$ is the reduced mass and
$a_{K^-p}$ the scattering length of the $K^-p$ system. From Refs.~\cite{Tovee:1971ga,Nowak:1978au},
we take decay ratios listed in Eqs.~(\ref{b})-(\ref{d}).\\[-2ex]
\begin{subequations}\label{atom}
\begin{align}\label{a}
\Delta E -i\Gamma/2&=(283\pm42)-i(271\pm55)~{\rm eV}\\
 \gamma&=\frac{\Gamma_{K^-p\rightarrow
 \pi^+\Sigma^-}}{\Gamma_{K^-p\rightarrow \pi^-\Sigma^+}} =2.38\pm0.04\label{b}\\
 R_n&=\frac{\Gamma_{K^-p\rightarrow
 \pi^0\Lambda}}{\Gamma_{K^-p\rightarrow
 \text{neutral}}}=0.189\pm0.015\label{c}\\
R_c&=\frac{\Gamma_{K^-p\rightarrow \pi^\pm\Sigma^\pm}}{\Gamma_{K^-p\rightarrow
\text{inelastic}}}=0.664\pm0.011\label{d}
\end{align}
\end{subequations}

The quantities listed in Eqs.~(\ref{atom}) are compared to the fit in Fig.~\ref{Atom}.

\begin{figure}[pt]
\begin{center}
\begin{tabular}{cc}
\hspace{-2mm}\includegraphics[width=0.25\textwidth]{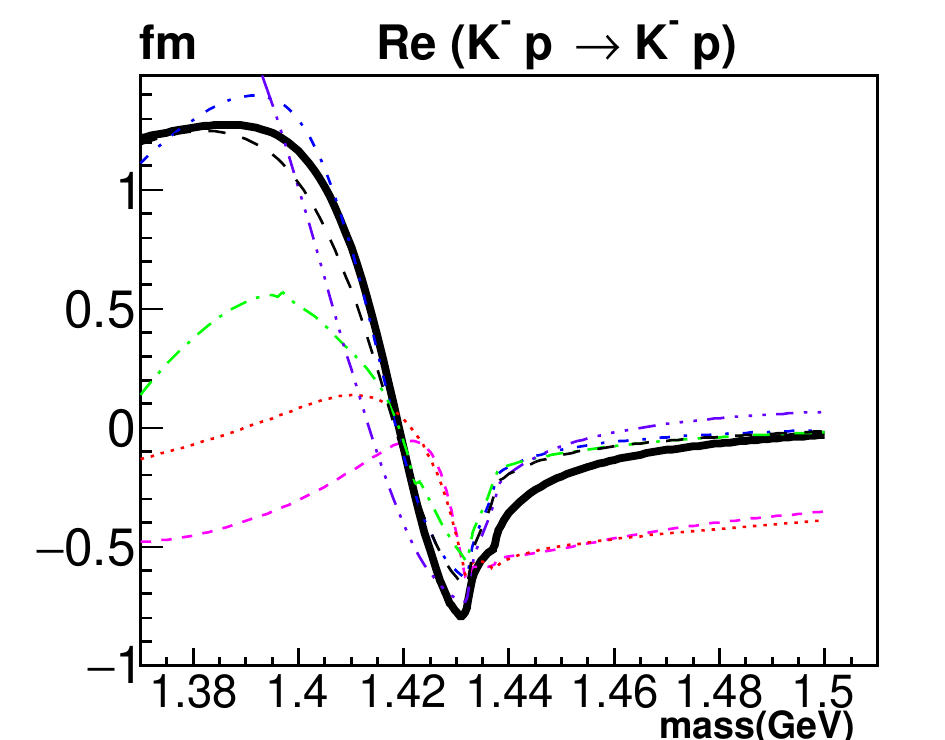}&
\hspace{-4mm}\includegraphics[width=0.25\textwidth]{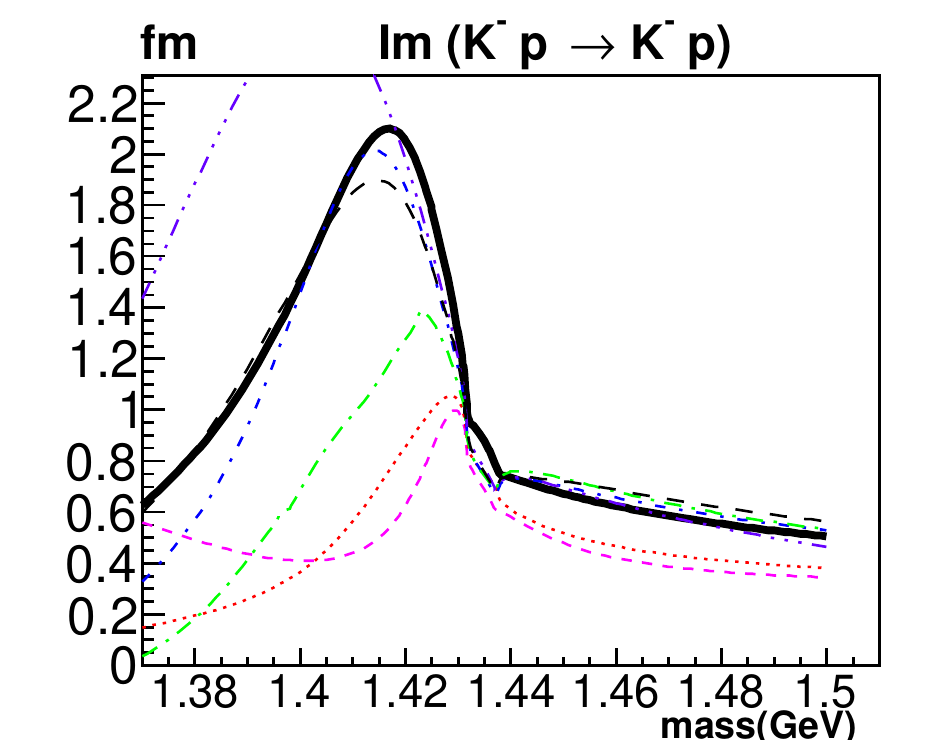}\\
\hspace{-2mm}\includegraphics[width=0.25\textwidth]{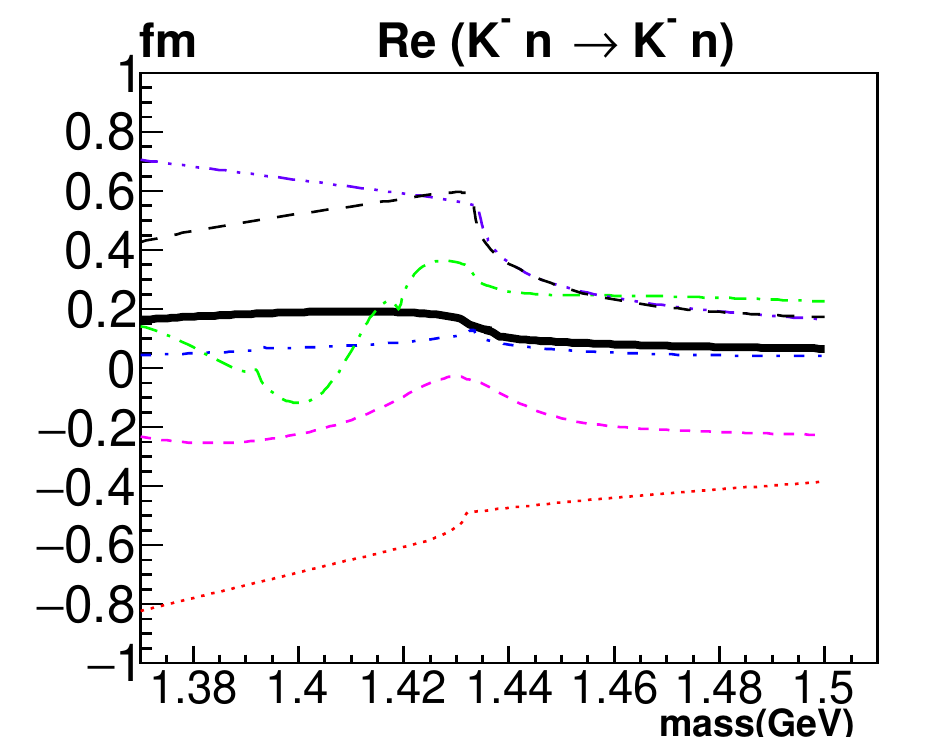}&
\hspace{-4mm}\includegraphics[width=0.25\textwidth]{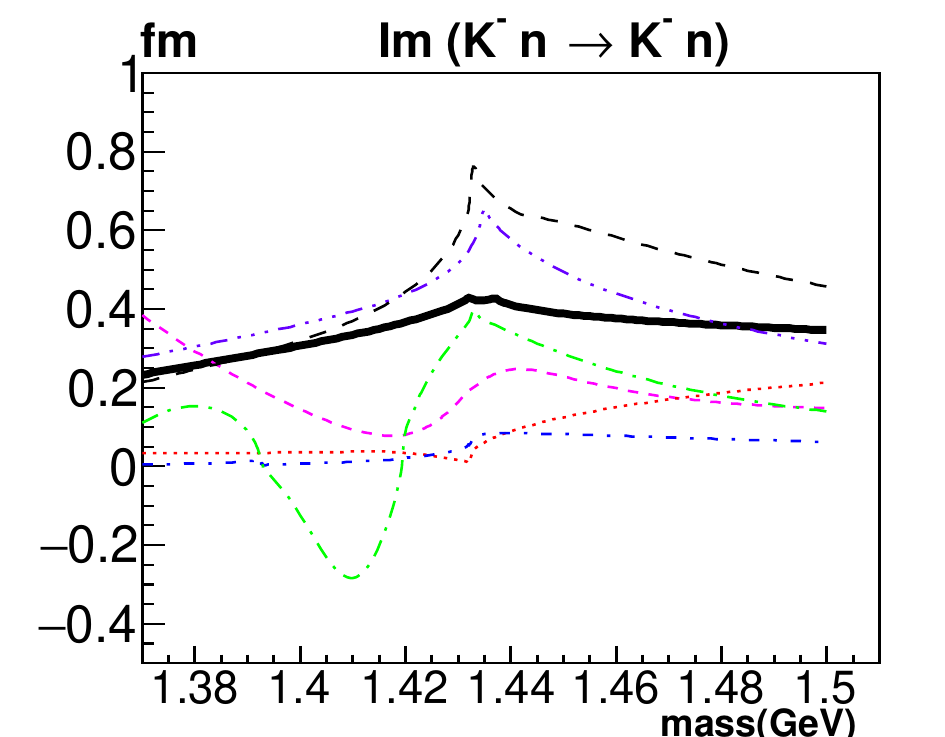}\\[-2ex]
\end{tabular}
\end{center}
\caption{\label{Scatt}$K^-p$ and $ K^- n$ S-wave elastic scattering amplitude. The black bold
curves corresponds to our one pole $\Lambda$ solution. A comparison with different other analyses 
is also given, see \cite{Cieply:2016jby} for details.
  }
\vspace{-2mm}
\end{figure}

The authors of Ref.~\cite{Cieply:2016jby} have performed a comparative analysis of the different
approaches based on the chiral SU(3) dynamics. The different approaches lead to rather different
predictions for the $K^-p$ and $K^-n$ S-wave elastic scattering amplitudes. In particular the
extrapolation to subthreshold energies yields a wide spectrum of results. The amplitudes are shown
in Fig.~\ref{Scatt} and compared to our $K^-p$ S-wave elastic scattering amplitudes. Our amplitudes
are well within the range of amplitudes derived in models based on the chiral SU(3) dynamics. The
real part of our scattering amplitude vanishes at about 1420\,MeV, the imaginary part reaches a
maximum of about 2\,fm.

\section{\label{Results}Results}

To find the pole positions in the $\Lambda(\frac12^-)$ wave in the region below 1500 MeV, we
performed one-pole and two-pole fits. The one-pole fit describes the data convincingly. The
two-poles hypothesis fit gives a slightly better description but we did not find any solution with
a pole position close to the $1300-1400$\,MeV region.  When a second pole
was admitted in the fit, it moved into
the non-physical region below the $\pi\Sigma$ threshold and can be considered as a non-resonant
background contribution; alternatively, the pole moved to the $K^-p$ threshold with an anomalously
small hadronic width (few MeV). We do not consider this solution as physically meaningful. In all
solutions, we find one leading pole position of the $\Lambda(1405)$. The pole properties are
collected in Table~\ref{results}.

The CLAS data on three-body final states
are, of course, more complicated to analyze; effects like three-body unitarity are not considered
in this analysis. When these data were excluded, the results hardly changed. In particular, no second pole
in the $1300-1400$\,MeV region was needed. The bubble chamber data from~\cite{Hemingway:1984pz}
had practically no impact on the fit; the data were included for historical reasons.

We use a two-pole parametrization for the $\Sigma(\frac12^-)$ partial wave. Both poles move far
away from the physical region and describe background processes, likely due to $t$ and/or
$u$-channel exchange processes. Thus we conclude that there is no $\Sigma(\frac12^-)$ resonance in
the region below 1500 MeV.

The transition residues listed  in Table~\ref{results} for the transition from the initial are defined as
\be
Res(K^-p\to \Lambda(1405)\to {\rm final}) = \frac{g_ig_f}{2W_{\rm pole}}\sqrt{\rho_i\rho_f}
\ee
where $W_{\rm pole}$ represents the pole mass and $\rho_i$, $\rho_f$ are the initial and final-state phase
spaces. The relative sign of the $K^-p\to \Lambda(1405)\to \pi\Sigma$ residue is cannot be deduced
from the available data.

\begin{table}[H]
\caption{\label{results} Residues for $K^-p\to \Lambda(1405)\to {\rm final}$.}
\renewcommand{\arraystretch}{1.4}
\bc
\begin{tabular}{lcc}
\hline\hline
Pole position: \qquad  $M_{\rm pole}$ = &\multicolumn{2}{c}{$(1422\pm 3, -i(21\pm3))$\,{\rm MeV}}\\
\hline
 Residues:                                                       &  Magnitude          & phase
                                                                    \\\hline
 $K^-p\to \Lambda(1405)\to \bar K N$            &  $63\pm 4$\,MeV  &  $(155\pm15)^\circ$\\
 $K^-p\to \Lambda(1405)\to \pi\Sigma$   &  $42\pm 3$\,MeV     &  $({0} \rm\,or\, {180}\pm 15)^\circ$\\
\hline\hline
\end{tabular}
\ec
\renewcommand{\arraystretch}{1.4}
\end{table}
\section{\label{Summary}Summary and discussion}

We have performed a partial wave analysis of low-energy data on $\Sigma\pi$ and $K^-p$ 
interactions. Analyses based on unitarized chiral perturbation
theory~\cite{Oller:2000fj,Jido:2003cb,Cieply:2009ea,Ikeda:2012au,Guo:2012vv,Mai:2012dt,Mai:2014xna,Roca:2013av,Roca:2013cca} find two $\Lambda^*$ resonances with $J^P=1/2^-$  in the region below 1500\,MeV.
We find that the data are fully compatible with a fit with one single resonance,
$\Lambda(1405)$, and background terms. The background consists of two or three poles below the
$\Sigma\pi$ threshold: two $\Sigma$ poles and either no or only one single $\Lambda$ pole. The pole
of the $\Lambda(1405)$ is found at $M_{\rm pole} = (1422\pm 3, -i(21\pm3))$\,{\rm MeV}.

\vspace{2mm}

One of the authors (UT) would like to acknowledge useful discussions with
U.-G. Mei{\ss}ner.
This work was supported by the Deutsche Forschungsgemeinschaft (SFB/ TR110), the
Russian Science Foundation (RSF 16-12-10267), the United States Department of Energy under contract DE-AC05-06OR23177 (Jefferson Lab) and DE-FG02-87ER 40315 (Carnegie Mellon University).


\begin{thebibliography}{99}
\bibitem{Alston:1961zzd}
  M.~H.~Alston {\it et al.}, 
  Phys.\ Rev.\ Lett.\  {\bf 6}, 698 (1961).
\bibitem{Dalitz:1960du}
  R.~H.~Dalitz and S.~F.~Tuan,
  Annals Phys.\  {\bf 10}, 307 (1960).
\bibitem{Dalitz:1967fp}
  R.~H.~Dalitz, T.~C.~Wong and G.~Rajasekaran,
  Phys.\ Rev.\  {\bf 153}, 1617 (1967).
\bibitem{Isgur:1978xj}
  N.~Isgur and G.~Karl,
  Phys.\ Rev.\ D {\bf 18}, 4187 (1978).
\bibitem{Kaiser:1996js}
  N.~Kaiser, T.~Waas and W.~Weise,
  Nucl.\ Phys.\ A {\bf 612}, 297 (1997).
\bibitem{Oller:2000fj}
  J.~A.~Oller and U.-G.~Mei\ss ner,
  Phys.\ Lett.\ B {\bf 500}, 263 (2001).
\bibitem{Jido:2003cb}
  D.~Jido, J.~A.~Oller, E.~Oset, A.~Ramos and U.-G.~Mei\ss ner,
  Nucl.\ Phys.\ A {\bf 725}, 181 (2003).
\bibitem{Oset:2001cn}
  E.~Oset, A.~Ramos and C.~Bennhold,
  Phys.\ Lett.\ B {\bf 527}, 99 (2002)
   Erratum: [Phys.\ Lett.\ B {\bf 530}, 260 (2002)]
  \bibitem{Cieply:2009ea}
  A.~Cieply and J.~Smejkal,
  Eur.\ Phys.\ J.\ A {\bf 43}, 191 (2010).
\bibitem{Ikeda:2012au}
  Y.~Ikeda, T.~Hyodo and W.~Weise,
  Nucl.\ Phys.\ A {\bf 881}, 98 (2012).
 \bibitem{Guo:2012vv}
  Z.~H.~Guo and J.~A.~Oller,
  Phys.\ Rev.\ C {\bf 87}, no. 3, 035202 (2013).
\bibitem{Mai:2012dt}
  M.~Mai and U.-G.~Mei\ss ner,
  Nucl.\ Phys.\ A {\bf 900}, 51  (2013).
%
\bibitem{Mai:2014xna}
  M.~Mai and U.-G.~Mei\ss ner,
  Eur.\ Phys.\ J.\ A {\bf 51},  30 (2015).
\bibitem{Roca:2013av}
  L.~Roca and E.~Oset,
  Phys.\ Rev.\ C {\bf 87},  055201 (2013).
\bibitem{Roca:2013cca}
  L.~Roca and E.~Oset,
  Phys.\ Rev.\ C {\bf 88}, 055206 (2013).
\bibitem{Miyahara:2018onh}
  K.~Miyahara, T.~Hyodo and W.~Weise,
  Phys.\ Rev.\ C {\bf 98}, 025201 (2018).
\bibitem{Feijoo:2018den}
  A.~Feijoo, V.~Magas and A.~Ramos,
  Phys.\ Rev.\ C {\bf 99}, no. 3, 035211 (2019).
\bibitem{Cieply:2016jby}
  A.~Cieply, M.~Mai, U.-G.~Mei\ss ner and J.~Smejkal,
  Nucl.\ Phys.\ A {\bf 954}, 17 (2016).
  \bibitem{Capstick:1986bm}
  S.~Capstick and N.~Isgur,
  Phys.\ Rev.\ D {\bf 34}, 2809 (1986)
  [AIP Conf.\ Proc.\  {\bf 132}, 267 (1985)].
\bibitem{Glozman:1995fu}
  L.~Y.~Glozman and D.~O.~Riska,
  Phys.\ Rept.\  {\bf 268}, 263 (1996).
    \bibitem{Loring:2001ky}
  U.~L\"oring, B.~C.~Metsch and H.~R.~Petry,
  Eur.\ Phys.\ J.\ A {\bf 10}, 447 (2001).
 \bibitem{Giannini:2015zia}
  M.~M.~Giannini and E.~Santopinto,
  Chin.\ J.\ Phys.\  {\bf 53}, 020301 (2015).
\bibitem{hyperon-II}
M.~Matveev, A.V.~Sarantsev, V.A. Nikonov, A.V. Anisovich, U. Thoma, and E.~Klempt, Hyperon II:
``Partial wave amplitudes for $K^-p$ scattering'', in preparation.
\bibitem{hyperon-III}
A.V.~Sarantsev, M.~Matveev, V.A. Nikonov, A.V. Anisovich, U. Thoma, and E.~Klempt, ``Hyperon III:
Properties of excited hyperons'', in preparation.
\bibitem{Aaij:2015tga}
  R.~Aaij {\it et al.} [LHCb Collaboration],
  Phys.\ Rev.\ Lett.\  {\bf 115}, 072001 (2015).
\bibitem{Fernandez-Ramirez:2015fbq}
  C.~Fernandez-Ramirez, I.~V.~Danilkin, V.~Mathieu and A.~P.~Szczepaniak,
  Phys.\ Rev.\ D {\bf 93}, no. 7, 074015 (2016).
\bibitem{Agakishiev:2012xk}
  G.~Agakishiev {\it et al.} [HADES Collaboration],
  Phys.\ Rev.\ C {\bf 87}, 025201 (2013).
 \bibitem{Hassanvand:2012dn}
  M.~Hassanvand, S.~Z.~Kalantari, Y.~Akaishi and T.~Yamazaki,
  Phys.\ Rev.\ C {\bf 87}, 055202 (2013).
\bibitem{Moriya:2013eb}
  K.~Moriya {\it et al.} [CLAS Collaboration],
  Phys.\ Rev.\ C {\bf 87}, 035206 (2013).
\bibitem{Moriya:2014kpv}
  K.~Moriya {\it et al.} [CLAS Collaboration],
  Phys.\ Rev.\ Lett.\  {\bf 112}, 082004 (2014).
\bibitem{Hassanvand:2017iif}
  M.~Hassanvand, Y.~Akaishi and T.~Yamazaki,
 ``Clear indication of a strong $I=0$  $\bar KN$ attraction in the $\Lambda (1405)$ region from the CLAS photo-production data,''
  arXiv:1704.08571 [nucl-th].
  \bibitem{Dong:2016auh}
  F.~Y.~Dong, B.~X.~Sun and J.~L.~Pang,
  Chin.\ Phys.\ C {\bf 41}, 074108 (2017).
\bibitem{Bruns:2010sv} 
  P.~C.~Bruns, M.~Mai and U.-G.~Mei{\ss}ner,
  Phys.\ Lett.\ B {\bf 697}, 254 (2011).
\bibitem{Myint:2018ypc}
  K.~S.~Myint, Y.~Akaishi, M.~Hassanvand and T.~Yamazaki,
  PTEP {\bf 2018}, 073D01 (2018).
 \bibitem{Lu:2013nza}
  H.~Y.~Lu {\it et al.} [CLAS Collaboration],
  Phys.\ Rev.\ C {\bf 88}, 045202 (2013).
  \bibitem{Prakhov:2004an}
  S.~Prakhov {\it et al.} [Crystall Ball Collaboration],
  Phys.\ Rev.\ C {\bf 70}, 034605 (2004).
  \bibitem{Hemingway:1984pz}
  R.~J.~Hemingway,
  Nucl.\ Phys.\ B {\bf 253}, 742 (1985).
\bibitem{Mast:1975pv}
T.~S.~Mast, M.~Alston-Garnjost, R.~O.~Bangerter, A.~S.~Barbaro-Galtieri, F.~T.~Solmitz and R.~D.~Tripp,
Phys.\ Rev.\ D {\bf 14}, 13 (1976).
\bibitem{Humphrey:1962zz}
  W.~E.~Humphrey and R.~R.~Ross,
  Phys.\ Rev.\  {\bf 127}, 1305 (1962).
\bibitem{Watson:1963zz}
  M.~B.~Watson, M.~Ferro-Luzzi and R.~D.~Tripp,
  Phys.\ Rev.\  {\bf 131}, 2248 (1963).
\bibitem{Sakitt:1965kh}
  M.~Sakitt, T.~B.~Day, R.~G.~Glasser, N.~Seeman, J.~H.~Friedman, W.~E.~Humphrey and R.~R.~Ross,
  Phys.\ Rev.\  {\bf 139}, B719 (1965).
\bibitem{Ciborowski:1982et}
  J.~Ciborowski {\it et al.},
  J.\ Phys.\ G {\bf 8}, 13 (1982).
  \bibitem{Tovee:1971ga}
  D.~N.~Tovee {\it et al.},
  Nucl.\ Phys.\ B {\bf 33}, 493 (1971).
\bibitem{Nowak:1978au}
  R.~J.~Nowak {\it et al.},
  Nucl.\ Phys.\ B {\bf 139}, 61 (1978).
\bibitem{Bazzi:2011zj}
  M.~Bazzi {\it et al.} [SIDDHARTA Collaboration],
  Phys.\ Lett.\ B {\bf 704}, 113 (2011).
\bibitem{Bazzi:2012eq}
  M.~Bazzi {\it et al.},
  Nucl.\ Phys.\ A {\bf 881}, 88 (2012).
\bibitem{Tanabashi:2018oca}
  M.~Tanabashi {\it et al.} [Particle Data Group],
  Phys.\ Rev.\ D {\bf 98}, no. 3, 030001 (2018).
\bibitem{Meissner:2004jr}
  U.-G.~Mei\ss ner, U.~Raha and A.~Rusetsky,
  Eur.\ Phys.\ J.\ C {\bf 35}, 349 (2004).
\bibitem{Gasser} G. Colangelo, J. Gasser, H. Leutwyler, Nucl. Phys.
 {\bf B603}, 125 (2001).
\bibitem{Caprini:2005zr}
  I.~Caprini, G.~Colangelo and H.~Leutwyler,
  Phys.\ Rev.\ Lett.\  {\bf 96}, 132001 (2006).
\bibitem{Salpeter} E. Salpeter and H.A. Bethe,
Phys. Rev. {\bf 84},
1232 (1951).
\end{thebibliography}
\end{document}